\documentclass[aps,prb,twocolumn,groupedaddress,showpacs,floatfix,superscriptaddress,longbibliography]{revtex4-2}
\usepackage[plainpages=false,pdfpagelabels,colorlinks=true,linkcolor=red,urlcolor=blue,citecolor=blue,pdftitle={Title},pdfauthor={},pdfdisplaydoctitle=true,pdfduplex=DuplexFlipLongEdge]{hyperref}
\usepackage{epsfig}
\usepackage{graphicx}
\usepackage[utf8]{inputenc}
\usepackage{amsmath,amssymb}
\usepackage{physics}
\usepackage[dvipsnames]{xcolor}

\usepackage[normalem]{ulem}

\begin{document}

\title{The Two Orbital, Interacting Hatano-Nelson Model}
\author{J. Huang}
\email{zchuang@ucdavis.edu}
\affiliation{Department of Physics and Astronomy, University of California, 
Davis, CA 95616, USA}
\author{N. Aggarwal}
\email{nqaggarwal@ucdavis.edu}
\affiliation{Department of Physics and Astronomy, University of California, 
Davis, CA 95616, USA}
\author{Rubem Mondaini}
\email{rmondaini@uh.edu}
\affiliation{Department of Physics, University of Houston, Houston, Texas 77004, USA}
\affiliation{Texas Center for Superconductivity, University of Houston, Houston, Texas 77204, USA}
\author{R.T. Scalettar}
\email{scalettar@physics.ucdavis.edu}
\affiliation{Department of Physics and Astronomy, University of California, 
Davis, CA 95616, USA}

\begin{abstract}
The single orbital, one-dimensional, Hatano-Nelson Hamiltonian
provides deep insight into the physics of non-Hermiticity, resulting from asymmetric left/right hopping, and its connections to localization. In the absence of disorder, its single particle eigenvalues  $E_{\alpha}$
lie on an ellipse in the complex plane whose extent in the imaginary direction is controlled by the degree of asymmetry.  When randomness is introduced, two sets of real eigenvalues emerge at the extremes of the largest and smallest 
real part of $E_{\alpha}$. These real eigenvalues are associated with localized eigenvectors. For spinless fermions, increasing near-neighbor interactions first
cause a transition to a charge density wave phase, and ultimately, on finite lattices, a collapse of all eigenvalues to the real axis. In this paper, we explore the presence of real eigenvalues in the interacting, two-particle sector for the spinful case (Hubbard model) in a two-chain (two-band) geometry
with a Hermitian interchain hopping. Our key results are to obtain the ``phase" diagrams for the existence of a purely real spectrum, as a function of the
interaction strength, degree of non-Hermiticity, and interchain hopping. We study the sensitivity to boundary conditions of the spectral properties of our two-chain model with winding number analysis and explore the relationship between PBC doublon states and OBC skin modes. To address the question of stability in such non-equilibrium systems, we solve the dynamics at low filling according to Lindbladian evolution and find that the non-Hermitian description is able to qualitatively describe such systems.
\end{abstract}


\maketitle

\section{Introduction}

The study of non-Hermitian Hamiltonians is acquiring increased interest~\cite{feinberg1999non,shen2018topological,ghatak2019new,Kawabata2019,ashida2020non,bergholtz2021exceptional,ding2022non,zhang2022review}. Bender~\cite{bender2007making} in particular has addressed the question of classes of continuous non-Hermitian Hamiltonians such as ${\cal \hat H} = \hat p^2 - \hat x^4$ which nevertheless retain a real spectrum, positive probabilities, and unitary time evolution, linking these properties to the existence of a continuous ${\cal PT}$ symmetry~\cite{feng2017non,el2018non}.   An especially active area of study is the connection of non-Hermiticity to open quantum systems~\cite{rotter2009non,song2019non,ashida2020non}.

The Hatano-Nelson Hamiltonian~\cite{hatano1996localization,hatano1997vortex,hatano1998non}, initially proposed as a description of flux line depinning in type-II superconductors, is one of the simplest models of non-Hermiticity.  Unlike continuum models, it is formulated on a (1D) {\it lattice}, with the non-Hermiticity arising from intersite hopping which is larger in one direction,  $t\,e^{+h}$,  than the other, $t\,e^{-h}$, a phenomenon which arises when a current introduces a Lorentz force which drives the flux lines preferentially~\footnote{In this paper, we will use a somewhat different convention for the hopping, namely $t\pm \delta$ in the two directions. See Eqs.~\ref{eq:H}-\ref{eq:U}}. Pinning is modeled by introducing random site energies. The basic phenomenology of the model is an eigenspectrum that lives on one-dimensional trajectories in the complex plane~\cite{feinberg1999spectral}. In the absence of disorder, this curve is an ellipse $\epsilon_k = t\,(e^{h + ik} + e^{-h-ik}) = 2t\,{\rm cosh}\,h \, {\rm cos}\,k + i\,2t\,{\rm sinh}\,h \, {\rm sin}\,k$, where ${\rm sinh}\,h$ is seen to control the ellipse axis length along the imaginary axis. The addition of disorder induces the formation of `wings' extending outward from the ellipse along the real axis.  The eigenfunctions associated with these real eigenvalues are localized, while the complex eigenvalues retain delocalized eigenfunctions. The eigenfunction localization length $\xi$ at the bifurcation point, where the real spectrum separates into the complex plane, takes the value $\xi=h^{-1}$~\cite{ledoussalunpub}.

In this work, we investigate the interplay between on-site electron-electron interactions and the coupling between two one-dimensional Hatano-Nelson chains,
thereby generalizing both the geometry of the Hatano-Nelson model
to several chains, and also to the presence of correlations arising from having two spin species and an on-site repulsion $U$, as is present in the Hubbard Hamiltonian. We obtain the complex-valued energy spectrum at fixed fillings using Exact Diagonalization (ED) and study the system's topology using winding-number calculations. The non-Hermitian systems we consider exhibit a complex-real transition in the eigenvalue spectrum under periodic boundary conditions (PBC), characterized by the winding number of the locus traced as the spectrum varies in the complex plane. We find that such a transition and its corresponding topology constitute a generic description of bulk systems with balanced asymmetry.

The presence of interchain hopping $V_0$ allows the interplay of non-Hermiticity and band structure, since, at half-filling, the two-chain model can be tuned through a metal-insulator transition by increasing $V_0$. In addition to providing theoretical insight into a ``phase" diagram delineating the regime of real spectra in an interacting non-Hermitian system, our work also offers new guidance for the interpretation of experiments on such realizations ~\cite{koh2025interacting}.

The remainder of this paper is organized as follows: In Sec.~\ref{sec:M&M}, we introduce our model, two Hatano-Nelson chains coupled by a Hermitian interchain hybridization, and also how its topology is probed by the winding number. Section~\ref{sec:results} presents our results, first in the single particle (non-interacting) limit, where we develop criteria for a real spectrum, and then in the presence of interactions in the sector of one up and one down electron. This discussion is followed by results for the winding number, spin modes, and Lindbladian dynamics. Finally, Sec.~\ref{sec:discussion} presents a discussion and summary of our key results.

\section{Model and Methods} \label{sec:M&M}

\subsection{The Hamiltonian}

\begin{figure}[t!]
    \includegraphics[width=1\columnwidth]{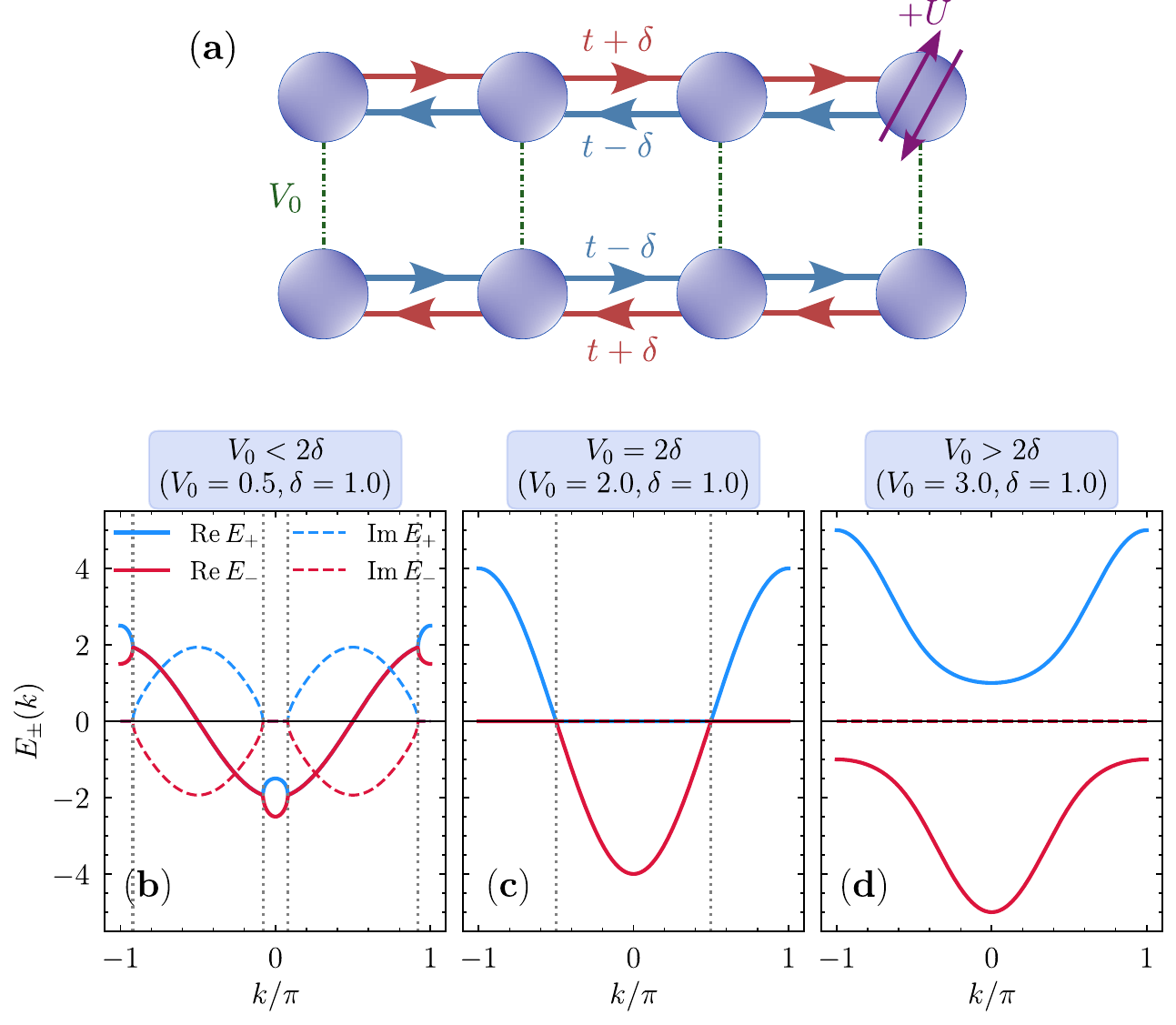}
    \caption{(a) Schematic representation of the Hamiltonian for the two-chain interacting Hatano-Nelson model with relevant terms annotated. Non-Hermitian hopping occurs along each of the chains $\alpha=A,B$ with hopping $t \pm \delta_\alpha$ for $j \rightleftarrows j+1$; the interchain hopping $V_0$ is Hermitian. Here, the sign of the non-Hermiticity is reversed among the two chains, $\delta_A=-\delta_B=\delta$. Occupation of up and down fermions on the same site leads to a correlation energy $U$. (b), (c), and (d) show the real and imaginary parts of the $E_\pm(k)$ bands for three representative values of the interchain hopping, with fixed $\delta = 1$, in the non-interacting limit, $U = 0$. The vertical dashed lines mark the exceptional points $k_{\rm EP}$ (see text).}
   \label{fig:fig_1}
\end{figure}

We consider the fermionic interacting Hatano-Nelson model~\cite{Zhang2022, Faugno2022, Dora2022, Kawabata2022, Alsallom2022, Longhi2023, Orito2023, Dupays2025} on two coupled rings, each of $L$ sites, 
\begin{equation}
    \hat{{\cal H}}=\hat{{\cal K}}+\hat{{\cal U}},
    \label{eq:H}
\end{equation}
where $\hat{\cal {K}}$ is the single-particle (`hopping') Hamiltonian,
\begin{eqnarray}
    \hat{{\cal K}}\hspace{-0.02in}=\hspace{-0.02in}-\hspace{-0.04in}\sum_{j,\alpha,\sigma} \hspace{-0.04in}&[&\hspace{-0.04in}(t+\delta^{\phantom{\dagger}}_\alpha) 
    \hat{c}_{j+1,\alpha \sigma }^\dag \hat{c}_{j,\alpha \sigma}^{\phantom{\dag}}
    +(t-\delta^{\phantom{\dagger}}_\alpha)
    \hat{c}^\dag_{j,\alpha \sigma} \hat{c}_{j+1,\alpha \sigma }^{\phantom{\dag}}]
    \nonumber \\
    -&V_0&\sum_{j,\sigma} (\hat{c}^\dag_{j,A\sigma}\hat{c}^{\phantom{\dagger}}_{j,B\sigma}
    +\hat{c}^\dag_{j,B\sigma}\hat{c}_{j,A\sigma}^{\phantom{\dag}})\ .
    \label{eq:K}
\end{eqnarray}
Here $\hat{c}_{j\alpha \sigma }^\dag (\hat{c}_{j\alpha \sigma}^{\phantom{\dag}})$ are creation (annihilation) operators for fermions on sites $j=1,2,\ldots, L$ and chains $\alpha=A,B$  with spin $\sigma$. The hopping along each chain $\alpha$ is non-Hermitian, taking values $t \pm \delta_\alpha$ for $j \rightleftarrows j+1$ and $\delta_A = -\delta_B \equiv \delta$, see Fig.~\ref{fig:fig_1}(a). We use periodic boundary conditions (PBC) along the chains unless stated otherwise. The interchain hopping $V_0$ is Hermitian, i.e., taking the same values for $A \leftrightarrow B$.
Meanwhile,
\begin{equation}
    \hat{\cal U}=U\sum_{j,\alpha} \hat{n}_{j,\alpha \uparrow}^{\phantom{\dagger}} 
    \hat{n}_{j,\alpha \downarrow}^{\phantom{\dagger}}
    \label{eq:U}
\end{equation}
is the usual Hubbard interaction term between the densities
$\hat n_{j,\alpha \sigma}^{\phantom{\dagger}} = \hat c_{j,\alpha \sigma}^{\dag}  \hat c_{j,\alpha \sigma}^{\phantom{\dag}}$ for spin $\sigma=\,\uparrow$ and $\sigma=\,\downarrow$ electrons on the same site $j$ at chain $\alpha$. In what follows, we focus on the two-particle sector with one fermion of each spin, $N_\uparrow=N_\downarrow=1$, which already captures the interplay between nonreciprocal hopping, interleg hybridization, and onsite interactions in its simplest nontrivial form. In addition, we establish $t=1$ as our energy scale.

Previous work on related interacting non-Hermitian lattice models has mainly followed three directions. First, several studies have examined higher-dimensional geometries in the non-interacting limit~\cite{zee1998a,Lee2019,li2020critical}. Second, interacting models with spin-dependent non-Hermiticity, $\delta_\uparrow=-\delta_\downarrow$, have been investigated on two-dimensional lattices using a combination of mean-field theory and quantum Monte Carlo, showing that non-Hermiticity suppresses antiferromagnetic order~\cite{Hayata2021, Yu2024}. Third, one-dimensional interacting variants~\cite{Zhang2022} have been considered both in the presence of spin-dependent asymmetry and spin-flip terms~\cite{Suthar2022}, and within dynamical mean-field theory, where correlations were found to reduce the skin effect~\cite{rangi2025interplay}. By contrast, the present work considers a ladder geometry with opposite nonreciprocity on the two legs and Hermitian interleg coupling, and focuses on the interacting two-particle problem.
Additional work has considered non-Hermitian hopping on a Bethe lattice~\cite{sun2025}.

\subsection{Computational Details}

Our numerical analysis is based on exact diagonalization of the Hamiltonian in Eq.~\eqref{eq:H} within the two-particle sector $N_{\uparrow}=N_{\downarrow}=1$. Since the two fermions carry opposite spin, they are distinguishable, and each can occupy any of the $2L$ sites of the ladder. The resulting Hilbert-space dimension is therefore $\dim {\cal H} = (2L)^2$. For PBC, we further exploit translational invariance along the ladder direction. The Hamiltonian then decomposes into $l=0,1,\dots,L-1$ momentum sectors, each of dimension $(2L)^2/L=4L$.

\begin{figure*}[t!]
    \includegraphics[width=1\textwidth]{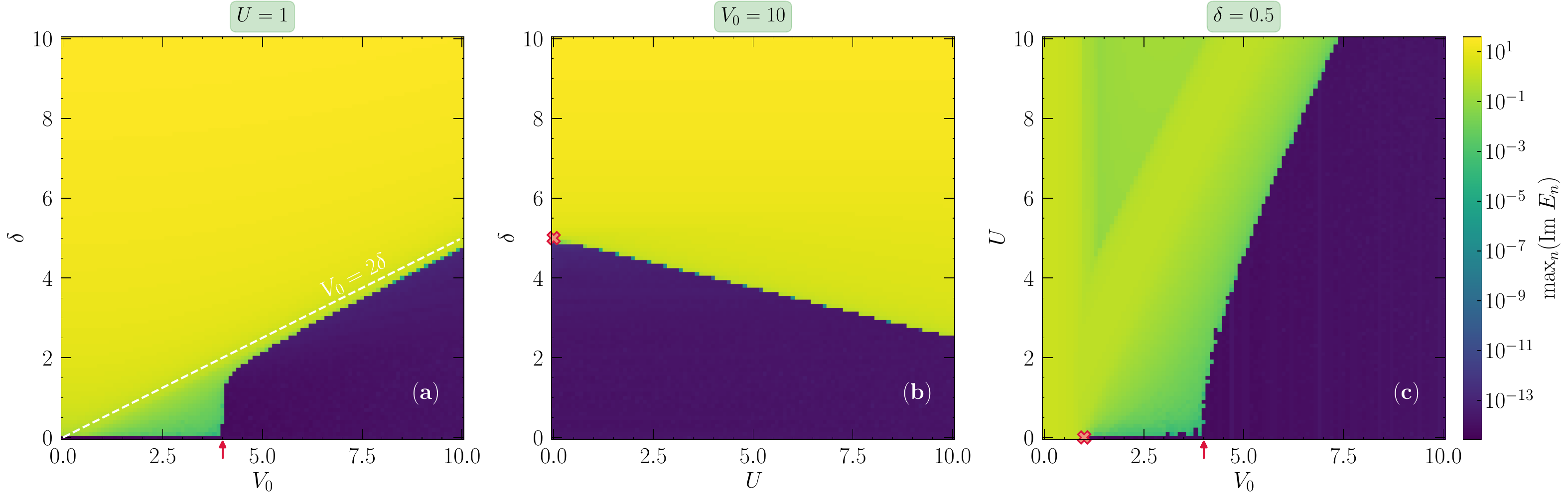}
    \caption{Two-dimensional heat maps of the largest imaginary part of the eigenspectrum for three sets of fixed parameters (a) $U =1$, (b) $V_0 = 10$, and (c) $\delta = 0.5$. The dashed line $V_0 = 2\delta$ in (a) marks the onset of a purely real spectrum in the non-interacting ($U = 0$) limit, corresponding to the exceptional-point condition discussed in Sec.~\ref{sec:singleparticle}. This can also be seen as the markers(red crosses) in panels (b) and (c), precisely in the $U=0$ axis. For finite $U$, one sees in panels (a) and (c) the many-body effects leading to a sharp increase in the critical value of $V_0$ where the real-complex transition occurs(red arrows). Data are extracted for an $L=100$ ladder with fixed spin-resolved particle numbers $N_\uparrow=N_\downarrow=1$. 
}
   \label{fig:fig_2}
\end{figure*}
A first question is whether the many-body spectrum is purely real or contains complex eigenvalues, which can be determined directly from the exact eigenspectrum. We also characterize the point-gap topology of the complex spectrum through a spectral winding number~\cite{Gong2018, Kawabata2019}, which is closely related to the non-Hermitian skin effect and to the modified bulk-boundary correspondence in non-Hermitian systems~\cite{Lee2016Ano, Xiong_2018, Borgnia2020, Zhang2020, Okuma2020}. To define the winding number, we thread magnetic fluxes through the two rings. Under the Peierls substitution~\cite{Peierls,Luttinger}, this amounts to adding phase factors to the intraleg hopping terms. The kinetic Hamiltonian becomes
\begin{align}
    \hat{\mathcal K}(\phi)
    = &-\sum_{j,\alpha,\sigma}
    \Big[
    e^{+i\phi_\alpha/L}(t+\delta_\alpha)\,
    \hat{c}_{j+1,\alpha\sigma}^\dag \hat{c}_{j,\alpha\sigma}
    \nonumber\\
    &\qquad\quad
    +\,e^{-i\phi_\alpha/L}(t-\delta_\alpha)\,
    \hat{c}_{j,\alpha\sigma}^\dag \hat{c}_{j+1,\alpha\sigma}^{\phantom{\dagger}}
    \Big]
    \nonumber\\
    &-V_0\sum_{j,\sigma}
    \left(
    \hat{c}_{j,A\sigma}^\dag \hat{c}_{j,B\sigma}
    +\hat{c}_{j,B\sigma}^\dag \hat{c}_{j,A\sigma}^{\phantom{\dagger}}
    \right)\ ,
\label{eq:Kphi}
\end{align}
while the interaction term $\hat{\mathcal U}$ remains unchanged. 

For a base energy $E$ lying within a point gap of $\hat{H}(\phi)$, a nontrivial topological invariant, the winding number, is defined as~\cite{Gong2018, Kawabata2019}
\begin{equation}
    W(E) = \oint_{0}^{2\pi} \frac{d\phi}{2\pi i}\,\frac{d}{d\phi}\ln \det\!\left[\hat{\cal H}(\phi)-E\right].
    \label{eq:wind1}
\end{equation}
This quantity counts how many times the complex quantity $\det[\hat{\cal H}(\phi)-E]$ winds around the origin as $\phi$ is varied from $0$ to $2\pi$. In a finite system, the eigenvalues of $\hat{H}(\phi)$ evolve continuously with $\phi$, tracing loops in the complex-energy plane under PBC. A nonzero value of $W(E)$ therefore signals a nontrivial point-gap topology of the many-body spectrum at the target energy $E$~\cite{Gong2018,Zhang2020,Okuma2020,Kawabata2022}.

\section{Results} \label{sec:results}

\subsection{Single Particle (non-interacting) limit} \label{sec:singleparticle}

In the non-interacting limit ($U=0$) with PBC, the energies can be computed analytically since the Hamiltonian is block-diagonal in momentum $k=k_m=\frac{2\pi m}{L}$ for integer $m$. The Hamiltonian decouples into the spin sectors with its projection onto the spin $\sigma$ sector $\hat {\cal H}_\sigma = \sum_k\hat c_{k,\sigma}^\dagger H_k^{\phantom{\dagger}} \hat c_{k,\sigma}^{\phantom{\dagger}}$, where
\begin{equation}
    H_k = \begin{pmatrix} \varepsilon_A(k) & -V_0 \\
                          -V_0 & \varepsilon_B(k)
    \end{pmatrix}\ .
\end{equation}
For the one-dimensional geometry of Fig.~\ref{fig:fig_1}, the decoupled intra-chain dispersion relation $\varepsilon_\alpha(k) = -(t + \delta_\alpha)e^{ik}-(t-\delta_\alpha)e^{-ik}$, with $\alpha = A,B$ and $\delta_A = -\delta_B \equiv \delta$. As such, the corresponding bands read~\cite{li2020critical}:
\begin{equation}
    E_{\pm}(k) = -2t\cos k \pm \sqrt{V_0^2-4\delta^2\sin^2 k}\ 
    \label{eq:dispersionreln}
\end{equation}

Equation \eqref{eq:dispersionreln} allows one to identify regions where the spectrum exhibits exceptional points, when eigenvalues and eigenvectors of the two bands coalesce~\cite{ding2022non}. These occur when $V_0^2-4\delta^2\sin^2 k = 0$, i.e., when $\sin k_{\rm EP} = \pm \frac{V_0}{2\delta}$, which admits up to four solutions in the first Brillouin zone, $k^{(1)}_{\rm EP} = \pm\arcsin\left(\frac{V_0}{2\delta}\right)$ and $k^{(2)}_{\rm EP} = \pm(\pi - \arcsin\left(\frac{V_0}{2\delta}\right))$, depending on the ratio $\frac{V_0}{2\delta}$.

We report the corresponding $E_{\pm}(k)$ bands in Fig.~\ref{fig:fig_1}(b--d), for three representative values of the interchain hopping $V_0$ at fixed $\delta=1$. For $V_0<2\delta$ [$V_0=0.5$, Fig.~\ref{fig:fig_1}(b)], the spectrum is partly complex, and the four exceptional points are located at the momenta where $V_0^2-4\delta^2\sin^2 k=0$. At these points, the two bands coalesce, while away from them the eigenvalues form complex-conjugate pairs over part of the Brillouin zone. At the critical value $V_0=2\delta$ [Fig.~\ref{fig:fig_1}(c)], the spectrum becomes entirely real, with a single pair of exceptional points at $k_{\rm EP}=\pm\arcsin(1)=\pm\pi/2$, marking the boundary between a regime in which part of the spectrum is complex and one in which all eigenvalues are real. For larger values of $V_0$ [$V_0=3$, Fig.~\ref{fig:fig_1}(d)], the spectrum remains entirely real. Since the Hamiltonian $\hat{\cal H}$ is $\mathcal{PT}$ symmetric throughout (Appendix~\ref{app:pt_symm}), this change is naturally interpreted as the transition from a $\mathcal{PT}$-broken regime, where some eigenstates occur in complex-conjugate pairs, to an unbroken regime, where the full spectrum is real.

These results emphasize that $V_0 \geq 2\delta$ sets a condition for obtaining a purely real spectrum for $U=0$, and it is our goal in what follows to understand how this condition survives when considering the interacting regime ($U \neq 0$) in the two-particle sector. In addition,  Appendix \ref{app:BC_spec} reviews the effect of boundary conditions for the single-chain Hatano-Nelson model and for the ladder model we investigate.

\subsection{Purely Real Spectra for $N_\uparrow = N_\downarrow=1$}
An equally simple analytical condition for the onset of this purely real regime in the $U\neq 0$ case does not exist; we therefore resort to numerical methods in the two-particle sector. For that, we diagonalize the Hamiltonian of Eq.~\eqref{eq:H} on ladders with $L=100$, sufficiently large to mitigate finite-size effects (see Appendix \ref{app:FSE}). We show the resulting ``phase'' diagrams in Fig.~\ref{fig:fig_2}, describing the largest imaginary part of the $\hat {\cal H}$ eigenspectrum, $\max_n({\rm Im}\ E_n)$, over different planes of the Hamiltonian parameters. We find that a finite $U$ value generally increases the imaginary part of the eigenvalues.  For example, in Fig.~\ref{fig:fig_2}(b), the emergence of complex eigenvalues occurs at smaller and smaller $\delta$ as $U$ increases.  

\begin{figure*}[t!]
\includegraphics[width=2\columnwidth]{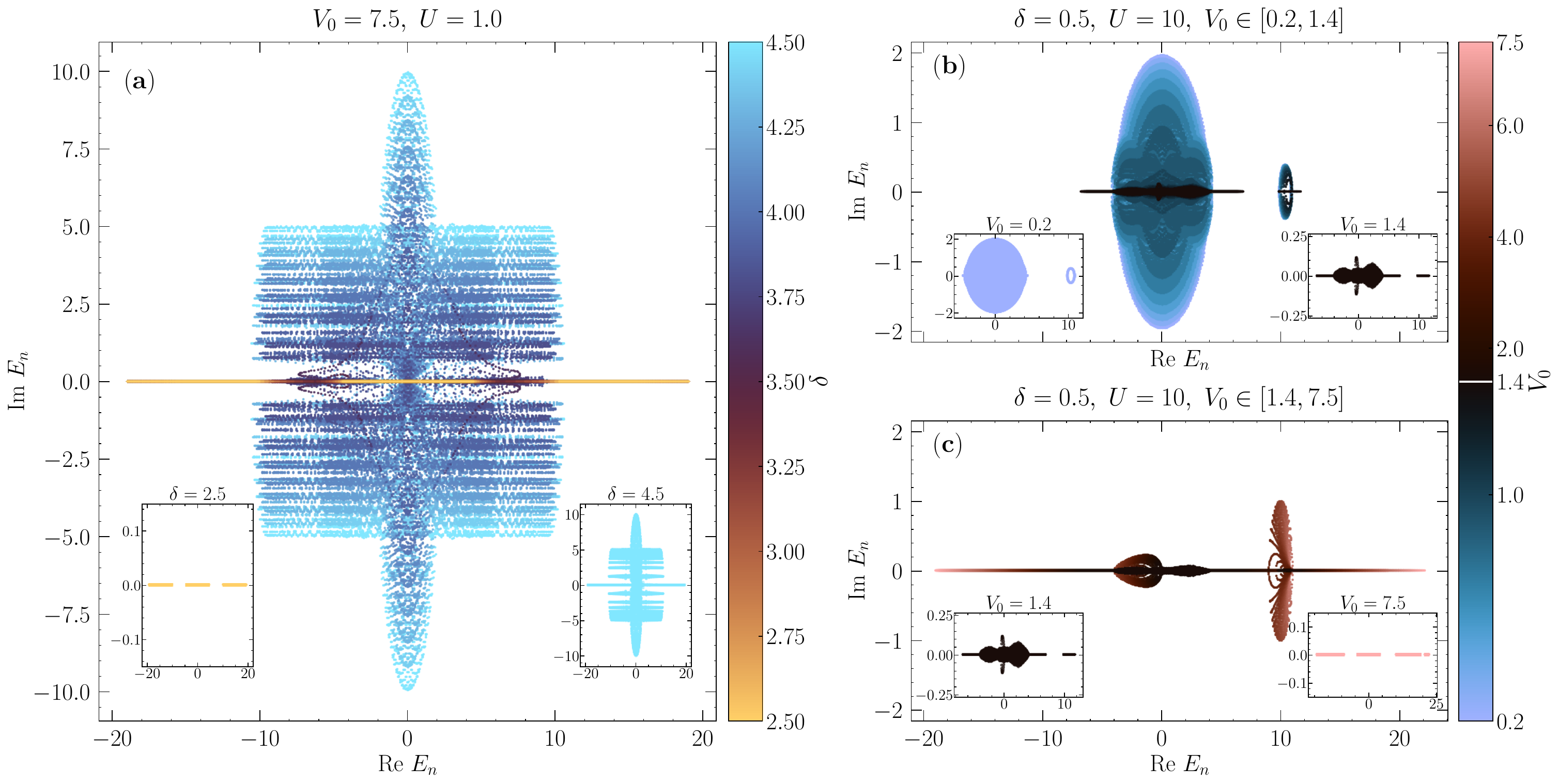}
    \caption{Representative spectra $E_n$ in the complex plane for the $N_\uparrow=N_\downarrow=1$ sector, illustrating how the eigenvalue distribution evolves across the parameter regimes discussed in Fig.~\ref{fig:fig_2}. (a) Fixed $V_0=7.5$ and $U=1$, with $\delta/t$ varying from $2.5$ to $4.5$. As $\delta$ increases, the spectrum moves away from the real axis and develops extended complex branches. (b),(c) Fixed $\delta=0.5$ and $U=10$, with $V_0$ varied over two ranges, shown separately for clarity: (b) $V_0\in[0.2,1.4]$ and (c) $V_0\in[1.4,7.5]$. In this case, the spectrum contains both a broad low-energy complex sector and a detached high-energy branch near ${\rm Re}\,E_n\sim U$, consistent with doublon-like states. Increasing $V_0$ progressively suppresses the imaginary extent of both sectors (even if non-monotonically) and eventually drives the full spectrum back onto the real axis. Insets show spectra at representative endpoint values. Points are colored by $\delta$ in (a) and by $V_0$ in (b),(c).}
    \label{fig:fig_3}
\end{figure*}

A useful reference point for understanding the weakly interacting regime is the Hermitian limit $\delta=0$, where the noninteracting two-particle spectrum consists of three scattering continua built from the bonding and antibonding single-particle bands. As discussed in Appendix~\ref{app:two_non_int_particles}, these continua become energetically separated only for $V_0>4$, providing a natural explanation for the characteristic scale $V_0\simeq 4$ associated with the onset of a purely real spectrum in Fig.~\ref{fig:fig_2}(a), and also visible in Fig.~\ref{fig:fig_2}(c) --- see vertical arrows. By contrast, in the regime of large $\delta$ and $V_0$, where $V_0,\delta\gg U$, the interacting spectrum may be viewed as a weakly perturbed version of the noninteracting ladder. In this regime, the numerical boundary for the onset of a purely real spectrum approaches the noninteracting condition $V_0\ge 2\delta$ [dashed line in Fig.~\ref{fig:fig_2}(a)].

To make the evolution of the spectrum more explicit, Fig.~\ref{fig:fig_3} shows the full set of eigenvalues $\{E_n\}$ in the complex plane along representative cuts of Fig.~\ref{fig:fig_2}. Figure~\ref{fig:fig_3}(a) corresponds to a vertical cut of Fig.~\ref{fig:fig_2}(a), at fixed $V_0=7.5$ and $U=1$, with increasing $\delta$. In this weakly interacting regime, the previously mentioned three real continua present at small $\delta$ progressively broaden and develop finite imaginary parts as $\delta$ approaches the noninteracting threshold $V_0/2$. 

In turn, the interplay between increasing $U$ and $V_0$ is subtle, since the limits in which either parameter is large while the other is small both tend to favor real eigenvalues. Building on the analytic result for $U=0$, a large but finite interaction generates an additional branch of eigenvalues with real part near $U$ and sizable imaginary component, consistent with doublon-like states and analogous to the Mott-Hubbard doublon branch in the single-chain case~\cite{Longhi2023}. While the principal branch associated with the remaining eigenvalues can be driven back to the real axis by increasing $V_0$, the full spectrum is not purely real unless the high-energy branch is also driven onto the real axis, which requires $V_0\gg U$.

This can be seen in Figs.~\ref{fig:fig_3}(b) and \ref{fig:fig_3}(c), corresponding to horizontal cuts of Fig.~\ref{fig:fig_2}(c) at fixed $\delta=0.5$ and $U=10$, as $V_0$ is increased. Even for a small $V_0$, the spectrum already contains a detached high-energy branch near ${\rm Re}\,E_n\sim U$, consistent with doublon-like states, in addition to a broad low-energy complex sector. As $V_0$ increases, the low-energy part of the spectrum is progressively pushed back toward the real axis. The high-energy branch, however, displays a non-monotonic evolution in its imaginary extent: it first narrows and becomes purely real near $V_0\simeq 1.4$, then acquires a finite imaginary support again at larger $V_0$, before eventually collapsing onto the real axis in the large-$V_0$ limit. Only for sufficiently large $V_0$ does the full spectrum become purely real.

\begin{figure}[t!]
    \includegraphics[width=1\columnwidth]{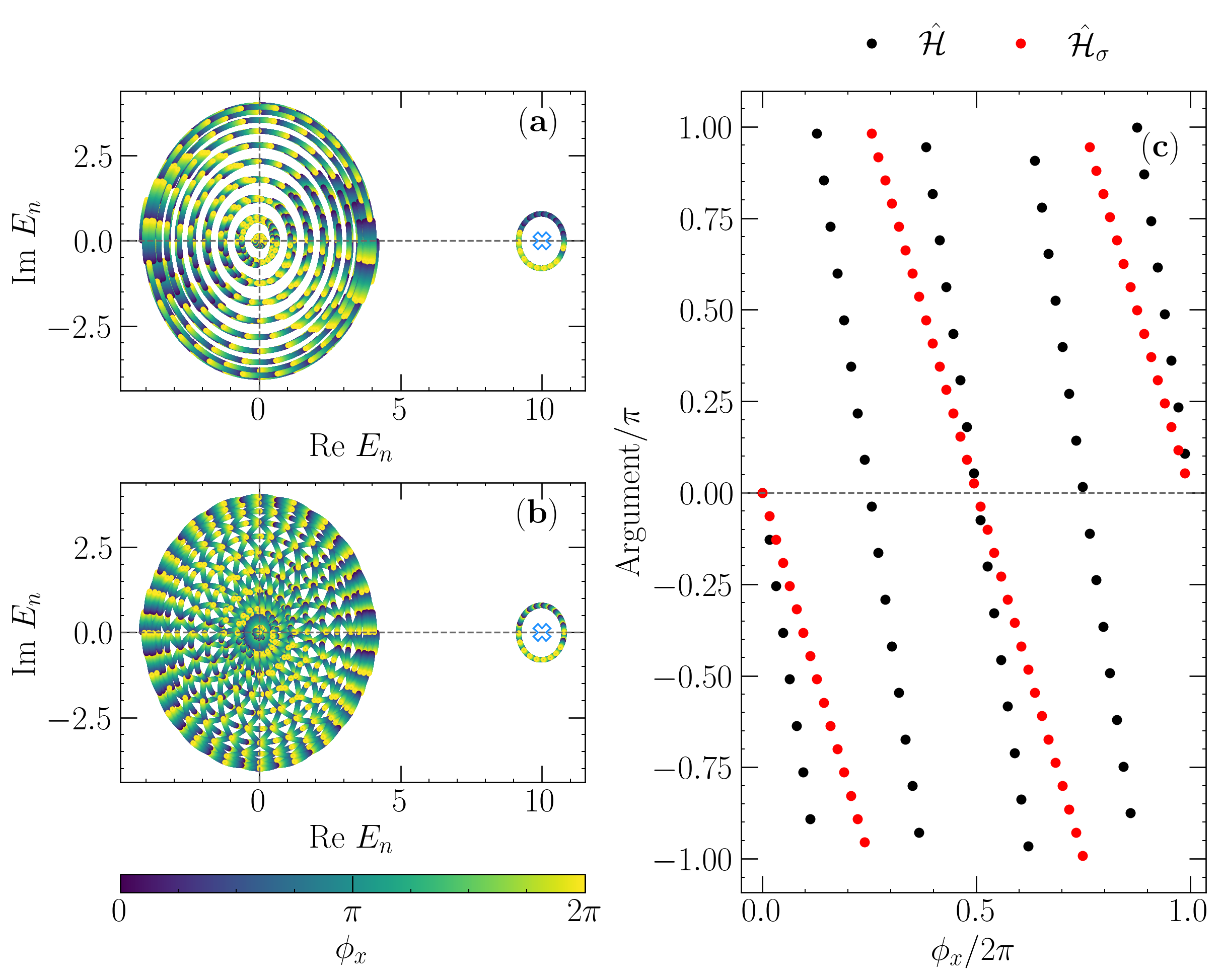}
    \caption{
    (a) Plot of the spectra in the complex plane under a continuously varying magnetic field. (b) Plot of the spectra as in (a), for the spin-resolved Hamiltonian. (c) Argument against flux $\phi$ for $E=U$ (plotted in black) and the spin-resolved Hamiltonian (red). Results are for the $N_\uparrow=N_\downarrow=1$ sector at fixed $U=10,$ $\delta=1$ and $V_0=0.1$ in an $L=20$ ladder. As $\phi$ varies from 0 to 2$\pi$, the discrete eigenvalues trace out concentric ellipses in the complex plane. The winding number $W=4$ is independent of the lattice size $L$ and Hamiltonian parameters as long as one remains in the regime where the spectrum is not fully real. Plots of the spectra and argument for $N_\uparrow=N_\downarrow=1$, $\delta=1$ and $V_0=0.1$. The spin-resolved winding numbers $W_{\sigma}=2$ for $\sigma=\uparrow,\downarrow$.}
   \label{fig:fig_4}
\end{figure}

\subsection{Winding Numbers}
The complex spectrum of the Hamiltonian considered here can exhibit nontrivial integer-valued winding numbers whenever there exists a reference energy $E$ inside a point gap of $\hat{\mathcal H}(\phi)$, namely when $\det[\hat{\mathcal H}(\phi)-E]\neq 0$ for all $\phi$ and the phase of this determinant winds as $\phi$ is varied~\cite{Gong2018, Kawabata2022}. In the strongly interacting regime, the detached high-energy doublon branch [Fig.~\ref{fig:fig_3}(b)] encloses such a point gap around $E\simeq U$. For the present ladder geometry with reversed nonreciprocity on the two legs, $\delta_A=-\delta_B\equiv\delta$, the appropriate probe is an opposite-leg flux pattern, $\phi_A=-\phi_B\equiv\phi$ in Eq.~\eqref{eq:Kphi}; otherwise, the resulting winding does not capture a nontrivial topological invariant (see Appendix~\ref{app:same_opp_flux}).

Figure~\ref{fig:fig_4}(a) shows the flux-resolved evolution of the eigenvalues of $\hat{\mathcal H}(\phi)$ for $\phi\in[0,2\pi)$. The low-energy sector consists of several overlapping elliptical loops and, for generic reference energies in that region, does not define an isolated point gap. By contrast, the detached doublon branch surrounds a clear point gap near $E\simeq U$. Choosing $E=U=10$, Fig.~\ref{fig:fig_4}(c) shows the corresponding variation of $\arg\det[\hat{\mathcal H}(\phi)-E]$, yielding the quantized winding number $W(E)=4$.

A natural question is whether this total winding can be decomposed into spin-resolved contributions. To probe this, we define a spin-selective flux insertion in which only one spin species acquires the opposite-leg flux, e.g.~
$\phi_{A,\sigma}=\phi$, $\phi_{B,\sigma}=-\phi$, while $\phi_{\alpha,\bar\sigma}=0$ for the other spin component. The corresponding spin-resolved winding number is then
\begin{equation}
    W_\sigma(E)=\oint_0^{2\pi}\frac{d\phi}{2\pi i}\,
    \frac{d}{d\phi}\ln\det\!\left[\hat{\mathcal H}_\sigma(\phi)-E\right],
    \label{eq:wind_sigma}
\end{equation}
where $\hat{\mathcal H}_\sigma(\phi)$ denotes the full interacting Hamiltonian with flux inserted only in the $\sigma$ sector. As shown in Fig.~\ref{fig:fig_4}(c), we find  $W_\uparrow(E)=W_\downarrow(E)=2$, so that the
total winding decomposes as $W(E)=W_\uparrow(E)+W_\downarrow(E)=4$ for the detached doublon point gap. Such additivity is not automatic in an interacting problem, but it does hold in the two-particle sector studied here.
The behavior at higher fillings is discussed in
Appendix~\ref{app:w_at_higher_densities}.

Note that in Fig.~\ref{fig:fig_4} we chose a weak interleg hybridization, $V_0=0.1$, so that the detached doublon branch encloses a clear point gap. At a larger $V_0$, this simple point-gap structure is lost. As discussed in Appendix~\ref{app:large_V0}, the flux-resolved spectra no longer exhibit the isolated doublon loop required for a robust winding-number assignment: depending on the interaction strength, the high-energy sector may either lose its detached character or deform into a more complicated set of structures. In either case, the relevant point gap closes, and the associated winding becomes trivial.

The importance of characterizing the winding numbers becomes clear when they are related to the corresponding skin modes, i.e., the accumulation of density at the edges under OBC (see Appendix~\ref{app:OBC} for a comparison of OBC and PBC spectra). This provides a direct connection between a topological property of the complex spectrum $\{E_n\}$ and a measurable spatial feature of the charge distribution~\cite{Mu2020,Lee2020,Alsallom2022,Zhang2022,Kawabata2022}. We explore this next.

\subsection{Skin Modes}

\begin{figure}[t!] 
\includegraphics[width=1\columnwidth]{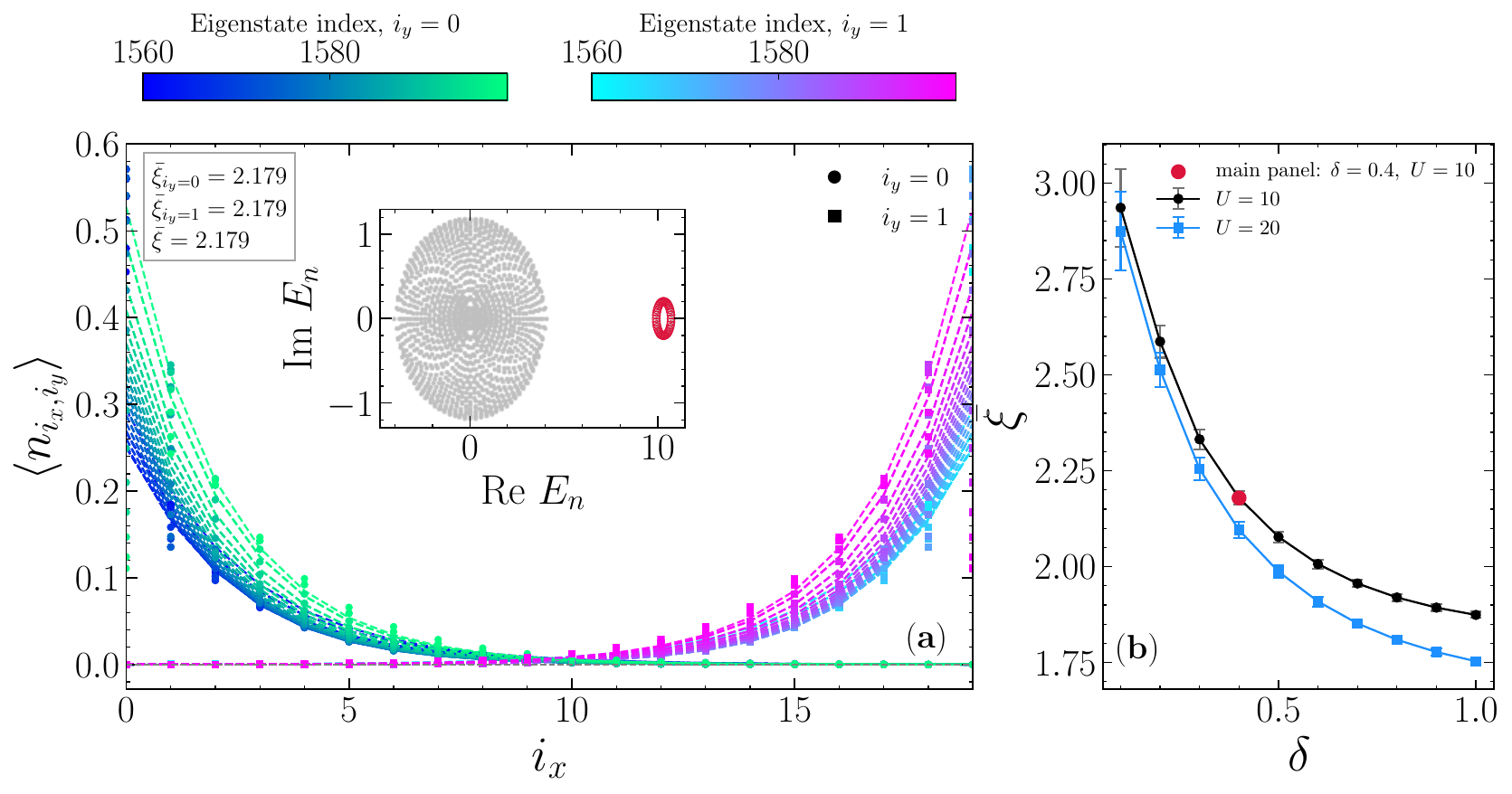}      \caption{(a) Density profiles under open boundary conditions for weak interchain hybridization $V_0=0.1$ in the $N_\uparrow=N_\downarrow=1$ sector, shown for the eigenstates highlighted in the inset and forming the doublon branch. The profiles are resolved for the lower chain ($i_y=0$, circles) and upper chain ($i_y=1$, squares). Dashed lines show exponential fits of the form $\propto e^{-x/\xi}$ for the lower chain and $\propto e^{-(L-1-x)/\xi}$ for the upper chain. (b) Average localization length $\bar{\xi}$ of the doublon branch as a function of the non-reciprocity parameter $\delta$, for both $U=10$ and $U=20$. The parameters are $L=20$, $V_0=0.1$, $U=10$, and $\delta=0.4$ in panel (a), while in panel (b) $L=20$ and $V_0=0.1$.} \label{fig:fig_5} \end{figure}

We characterize the skin effect under OBC by analyzing the spatial profiles of right eigenstates belonging to the detached doublon branch. For a normalized right eigenstate $|\psi_R\rangle$ of $\hat {\cal H}$, we evaluate the local density
\begin{equation}
    \langle \hat n_{i_x,i_y} \rangle_{\psi_R} =\sum_\sigma
    \langle \psi_R|\hat{c}^\dagger_{(i_x,i_y),\sigma}
    \hat{c}_{(i_x,i_y),\sigma}^{\phantom{\dagger}}| \psi_R \rangle .
\end{equation}
Figure~\ref{fig:fig_5}(a) shows the resulting density profiles for representative right eigenstates in the high-energy branch with ${\rm Re}\,E_n\approx U$. The densities decay exponentially along the ladder, as expected for skin modes, but because the nonreciprocity is opposite on the two legs, the accumulation occurs at opposite edges:
\begin{equation}
    \langle \hat n_{i_x,0} \rangle \propto e^{-i_x/\xi},
    \qquad
    \langle \hat n_{i_x,1} \rangle \propto e^{-(L-1-i_x)/\xi},
\end{equation}
where $\xi$ is the skin localization length. By fitting the profiles of all states in the detached branch to these functional forms, we extract $\xi$ for each state and then define their average, $\bar\xi$. The results are summarized in Fig.~\ref{fig:fig_5}(b).

We find that $\bar\xi$ decreases monotonically with increasing nonreciprocity $\delta$, and that for fixed $\delta$ the branch is more localized at larger interaction strength $U$. The first trend is consistent with the stronger nonreciprocal bias expected from the single-chain Hatano-Nelson limit. The second indicates that interactions further localize the detached doublon-like branch, consistent with the reduced spatial extent of strongly bound pairs in this high-energy regime. Thus, both increasing $\delta$ and increasing $U$ enhance the skin localization of the doublon sector.

\subsection{Lindbladian dynamics} \label{sec:lindbladian}

The existence of skin modes and their localization in a non-Hermitian Hamiltonian should not obscure the fact that such structures can arise naturally in open quantum systems. In particular, within the quantum-trajectory formulation of Markovian dynamics, the conditional no-jump evolution is governed by a non-Hermitian effective Hamiltonian, which is often a good approximation to the short-time dynamics. For seeing this, consider the density matrix $\hat\rho$, which obeys the Lindblad master equation ($\hbar\equiv 1$)
\begin{equation}
    \frac{d\hat\rho}{dt} = -i[\hat H_0,\hat\rho]
    +\sum_{j,\alpha,\sigma} \left(
    \hat L_{j,\alpha\sigma}\hat\rho \hat L^\dagger_{j,\alpha\sigma}
    -\frac12
    \left\{
    \hat L^\dagger_{j,\alpha\sigma}\hat L_{j,\alpha\sigma},\hat\rho
    \right\}
    \right)\ ,
    \label{eq:master_eq}
\end{equation}
where $\hat H_0$ is the Hermitian part of the Hamiltonian $\hat {\cal H}$, i.e.,
\begin{align}
        \hat{H}_0=&-t\sum_{j,\alpha,\sigma}      \left(\hat{c}_{j+1,\alpha \sigma }^\dagger \hat{c}_{j,\alpha \sigma}^{\phantom{\dagger}}
    +   \hat{c}^\dagger_{j,\alpha \sigma} \hat{c}_{j+1,\alpha \sigma }^{\phantom{\dagger}}\right)
    \nonumber \\
    &-V_0\sum_{j,\sigma} \left(\hat{c}^\dagger_{j,A\sigma}\hat{c}^{\phantom{\dagger}}_{j,B\sigma}
    +\hat{c}^\dagger_{j,B\sigma}\hat{c}_{j,A\sigma}^{\phantom{\dagger}}\right) \notag \\
    &+ U\sum_{j,\alpha} \hat{n}_{j,\alpha \uparrow}^{\phantom{\dag}} 
    \hat{n}_{j,\alpha \downarrow}^{\phantom{\dag}}\ .
\end{align}
If neglecting the jump terms, $\hat L_{j,\alpha\sigma}\hat\rho \hat L^\dagger_{j,\alpha\sigma}$, Eq.~\eqref{eq:master_eq} corresponds to the `no-jump evolution'
\begin{equation}
    \frac{d\hat\rho}{dt} = -i\left( \hat H_{\rm eff}\hat\rho-\hat\rho \hat H_{\rm eff}^\dagger \right)\ ,
\end{equation}
with
\begin{equation}
    \hat H_{\rm eff} = \hat H_0 -\frac{i}{2}\sum_{j,\alpha,\sigma}
    \hat L^\dagger_{j,\alpha\sigma}\hat L_{j,\alpha\sigma}\ .
\end{equation}

For the specific microscopic connection to the bath, we choose
\begin{align}
    \hat L_{j,0\sigma}&= \sqrt{2\delta}\,
    (\hat c_{j+1,0\sigma}-i\hat c_{j,0\sigma})\ ,
    \notag\\
    \hat L_{j,1\sigma}&= \sqrt{2\delta}\, (\hat c_{j-1,1\sigma}-i\hat c_{j,1\sigma})\ ,
    \label{eq:HN_Lindblad}
\end{align}
which generate opposite nonreciprocal hoppings on the two legs in $\hat H_{\rm eff}$. Indeed, the effective Hamiltonian in this regime reproduces the interacting Hatano-Nelson ladder introduced in Eq.~\eqref{eq:H}, with $\delta_A=-\delta_B\equiv\delta$, up to a uniform imaginary shift
\begin{equation}
\hat{H}_{\rm eff} = \hat{\cal H}-2i\delta\,\hat N\ ,
\end{equation}
where $\hat N=\sum_{j,\alpha,\sigma}\hat n_{j,\alpha\sigma}$ is the total number operator~\footnote{For OBC, the no-jump effective Hamiltonian generated by Eq.~\eqref{eq:HN_Lindblad} acquires a nonuniform imaginary onsite term, since edge sites participate in fewer jump operators than bulk sites. This can be remedied by introducing additional boundary jump operators, equivalently viewed as couplings to fictitious vacuum sites outside the chain. With this completion, the OBC effective Hamiltonian again reproduces the Hatano-Nelson ladder up to the uniform shift $-2i\delta\,\hat N$. We stress, however, that these boundary jump operators do affect the full Lindblad dynamics, even though they merely restore a homogeneous imaginary potential in the no-jump sector.}. Within a fixed-particle-number sector, this shift is an overall constant and therefore does not affect the eigenstates or the skin-mode structure. In real-time evolution, it only multiplies the wavefunction by a global decay factor; once the state is normalized, this factor drops out of all normalized observables. In the full Lindblad dynamics, however, the jump operators do not conserve particle number, so the system generically evolves toward sectors with fewer particles and ultimately toward the vacuum state.

As a first approximation, we therefore consider the dynamics generated solely by the effective non-Hermitian Hamiltonian, which conserves both the total particle number and each spin population. In this conditional no-jump regime, determining the conditions under which the spectrum is purely real remains meaningful, since the uniform imaginary shift $-2i\delta\,\hat N$ affects only the overall norm of the state and not the relative evolution within a fixed-particle-number sector. Thus, after normalization,
\begin{equation}
    |\psi(\tau+d\tau)\rangle= \frac{e^{-i\hat H_{\rm eff}d\tau}|\psi(\tau)\rangle} {\|e^{-i\hat H_{\rm eff}d\tau}|\psi(\tau)\rangle\|}\ ,
\end{equation}
the remaining spectral structure governs the transient dynamics. This approximate treatment is similar in spirit to that employed in Ref.~\cite{Longhi2023} for the single-chain case.

\begin{figure}[t!]
    \includegraphics[width=1\columnwidth]{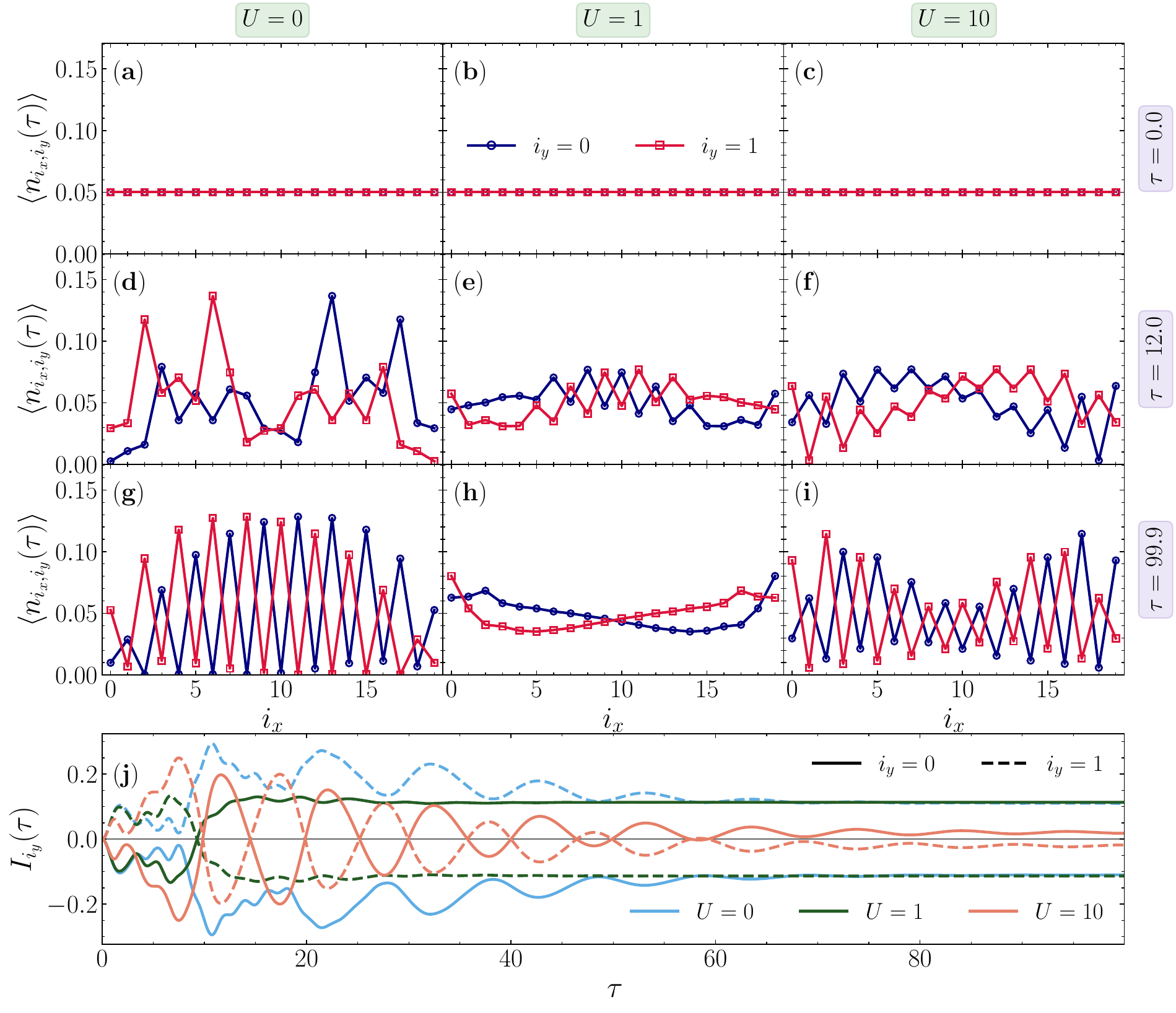}
    \caption{Deterministic no-jump dynamics of the Hatano-Nelson ladder for $L_x=20$, $V_0=1$, $\delta=0.5$. (a)-(i) show the normalized site-resolved densities $\langle n_{i_x,i_y}(\tau)\rangle$ along the two legs for three particular times and interaction strengths as indicated. (j) shows the leg-resolved imbalance. Because no quantum jumps are applied, the particle number is conserved. Time scales $\tau$ are plotted in units of $1/t$.}
    \label{fig:fig_6}
\end{figure}

As an illustrative example, we consider the dynamics from an initial state in the sector $N_\uparrow=N_\downarrow=1$, consisting of a uniform single-particle superposition on each leg,
\begin{equation}
    |\Psi(0)\rangle = \frac{1}{L} \Big(\sum_{i=1}^L \hat{c}_{i, A\uparrow}^\dagger\Big) \Big(\sum_{j=1}^L \hat{c}_{j,B\downarrow}^\dagger\Big)|0\rangle\ ,
\label{eq:psi0}
\end{equation}
where $|0\rangle$ is the vacuum state. 
Because the hopping bias is reversed between the two legs, one might naively expect opposite-edge accumulation on the two layers. However, the initial state $|\Psi(0)\rangle$ has broad overlap with the spectrum and does not selectively populate the detached doublon branch whose eigenstates display the clearest skin localization. As a result, the normalized no-jump dynamics does not generically reproduce a clean skin-mode profile. Instead, depending on $U$, one observes either broad layer-resolved edge accumulation or more oscillatory density patterns, as shown in Fig.~\ref{fig:fig_6}(a)--(i). This behavior is summarized by the leg-resolved imbalance $I_{i_y}(\tau)$, 
\begin{equation}
    I_{i_y}(\tau)=\sum_{i_x\in \mathrm{left\ half}} n_{i_x,i_y}(\tau)-\sum_{i_x\in \mathrm{right\ half}} n_{i_x,i_y}(\tau)\ ,
\end{equation}
where positive (negative) values indicate greater weight on the left (right) half of the ladder. Its dynamics for three interaction strengths is shown in Fig.~\ref{fig:fig_6}(j). While the imbalance persists at long times, no clear pattern associated with specific skin modes following the non-reciprocity direction is seen: The long-time imbalance for no- and strong interactions ($U=0$ and $U=10$) shows the same layer-resolved signatures, i.e., accumulation of charges in the right (left) parts of the ladder for the lower (upper) chain. On the other hand, small interactions ($U=1$) reverse this picture. This points out that this type of deterministic, yet non-unitary dynamics is not exactly suitable to generically observe the expected skin modes.

\begin{figure}[t!]
    \includegraphics[width=1\columnwidth]{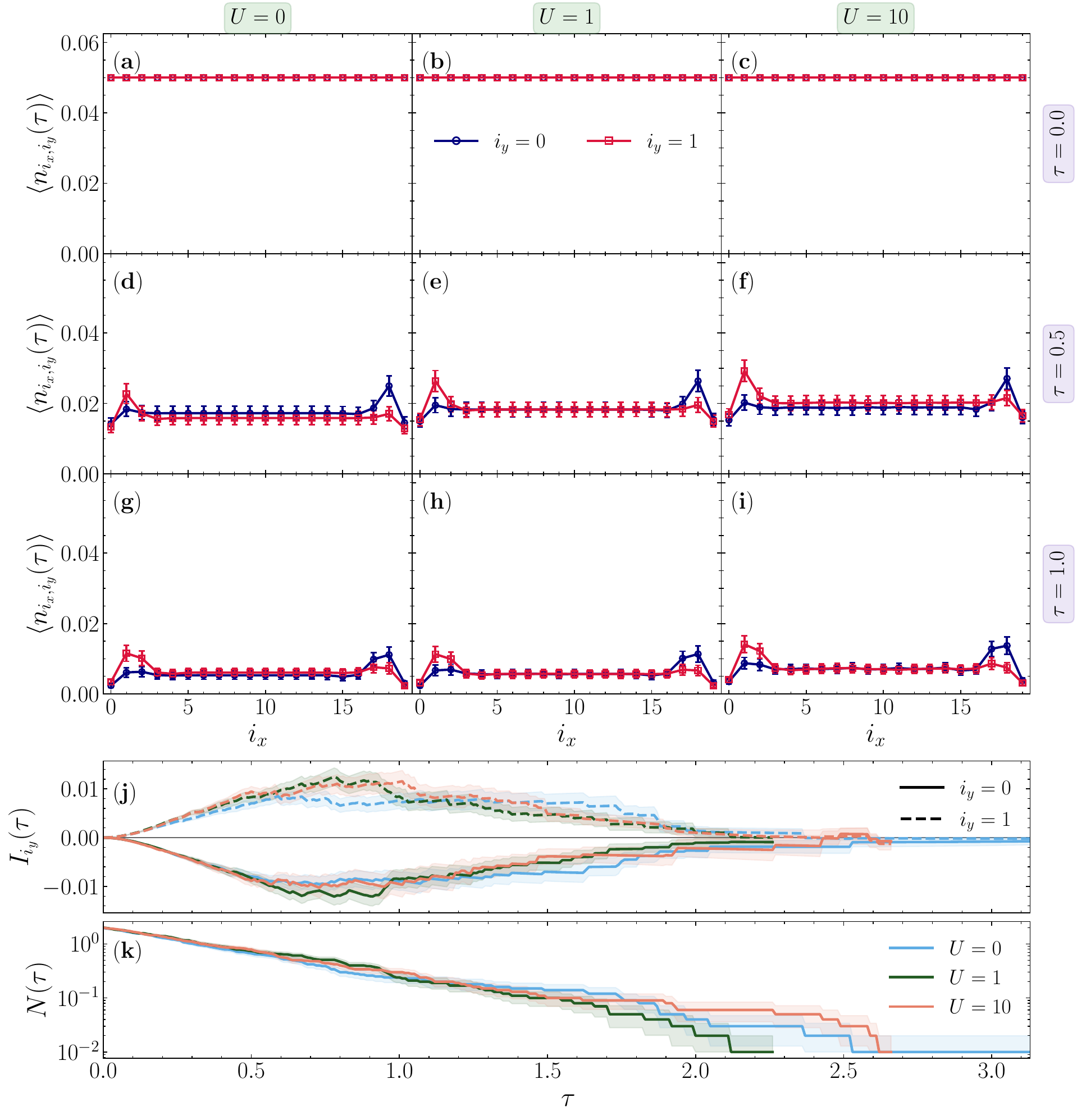}    
    \caption{Dynamics of the coupled two-chain system obtained from solving the {\it full} time-dependent Lindblad master equation including jump operator dissipation using the quantum trajectory method. Panels (a)-(i) show the site-resolved densities $\langle n_{i_x,i_y}(\tau)\rangle$ along the two legs, for $\tau=0.0,\,0.5,\,1.0$ and interaction strengths $U=0,\,1,\,10$. Panel (j) shows the leg-resolved imbalance $I_{i_y}(\tau)$ (see text), where solid and dashed lines correspond to $i_y=0$ and $i_y=1$, respectively. (k) Total particle number $N(\tau)$ on a logarithmic scale. In all panels, error bars or shaded regions denote the standard error of the mean over $100$ quantum trajectories. Parameters are $L=20$, $V_0=1$ and $\delta = 0.5$.
    }
    \label{fig:fig_7}
\end{figure}

The situation changes when considering the full dynamics of the quantum master equation, making use of the jump terms in the quantum trajectory method~\cite{dalibard1992wave,molmer1993monte,dum1992spe,dum1992vtsr,Daley2014}. Details of our implementation can be found in Refs.~\cite{Wang2023, Yi2025}. In this case, we find that the skin effect survives on time scales comparable to the system lifetime defined by dissipation effects when one considers the full Lindbladian dynamics by including the jump terms for different interaction strengths, see Fig.~\ref{fig:fig_7}(a)--(i). The imbalance [Fig.~\ref{fig:fig_7}(j)], unlike in the case of the no-jump dynamics, builds up with time but eventually returns to zero. The reason is not a redistribution of the charges, but rather because the system starts losing particles to the bath, ultimately reaching the vacuum state over fairly short time scales $\tau \gtrsim 1/t$. The decay of the total particle number $N(\tau) = \langle \hat N(\tau)\rangle$ is shown in Fig.~\ref{fig:fig_7}(k), displaying a characteristic exponential decay, and is fairly independent of the interaction strength.

\section{Summary and Discussion} \label{sec:discussion}
We have investigated an interacting two-leg Hatano-Nelson ladder in which the hopping nonreciprocity is reversed between the two legs and the two chains are coupled by a Hermitian interleg hybridization within the dilute sector $N_\uparrow=N_\downarrow=1$. In the noninteracting limit, the periodic-boundary spectrum becomes entirely real once the interleg coupling reaches the exceptional-point threshold $V_0=2\delta$. Once onsite repulsion is included, this simple condition is modified substantially: in the weakly interacting regime the onset of a purely real spectrum shifts to a scale $V_0 \simeq 4t$, consistent with the separation of the Hermitian two-particle continua, whereas at stronger coupling a detached high-energy branch with doublon-like character develops near ${\rm Re}\,E \sim U$, so that the spectrum is driven back to the real axis only once the hybridization becomes sufficiently large compared with the interaction scale.

The overall picture is therefore not one in which interleg hybridization merely suppresses non-Hermiticity in a monotonic fashion. Rather, hybridization, interactions, and the underlying two-particle continua compete. In particular, finite-size scaling (Appendix~\ref{app:FSE}) supports the stability of the main spectral trends, and the comparison between periodic and open boundaries (Appendix~\ref{app:OBC}) shows that, away from the decoupled limit $V_0=0$, the real-complex crossover remains qualitatively similar under both boundary conditions. In this sense, the emergence of real eigenvalues in the ladder is not tied solely to the familiar single-chain similarity transformation under OBC, but rather reflects the broader role of balanced nonreciprocity together with interleg coupling. Finally, calculations at higher density (Appendix~\ref{app:Many_Body}) and in a simple multilayer extension (Appendix~\ref{app:Bulk}) point in the same qualitative direction, although limitations of small-system exact diagonalization are more evident.

In the regime where the doublon branch remains spectrally isolated, it also supports a simple point-gap topology. Using a flux insertion pattern chosen to follow the opposite nonreciprocity on the two legs, we obtained a quantized winding number $W=4$, which in the dilute sector decomposes into spin-resolved contributions $W_\uparrow=W_\downarrow=2$. Under open boundary conditions, the corresponding states acquire clear skin-mode character, with charge accumulation at opposite edges of the two legs and a localization length that decreases as either $\delta$ or $U$ is increased.

A further result is the comparison between the effective non-Hermitian description and the dynamics of the full open system. In the normalized no-jump evolution (previously used in the literature~\cite{Longhi2023}), broad initial states do not, in general, resolve the clean-skin profiles associated with the isolated doublon sector. By contrast, the full Lindblad dynamics does display transient edge accumulation on time scales comparable to the dissipative lifetime before particle loss eventually drives the system toward the vacuum. Taken together, these results suggest that the spectral and topological features of the ladder are not merely formal properties of the effective Hamiltonian, but can leave observable dynamical signatures within a finite time window.

As a whole, our study paves the way for further investigation into the interplay among coupling, interactions, boundary conditions, the reality of the eigenspectrum, and the physical stability of non-Hermitian phenomena. 

We would like to conclude by indicating some connections of the present work across
disciplines. Although we have framed our discussion predominantly in the language of itinerant electrons and their interactions, it is worth emphasizing that our analysis and conclusions apply more broadly. In particular, the Hatano-Nelson model, and its topologically non-trivial windings, have recently been realized in the `synthetic dimension' formed by optical frequency modes in a modulated ring-resonator~\cite{wang2021generating}. Proposals have been made to include on-site `Hubbard' interactions, such as those studied here via the non-linear susceptibility~\cite{yuan2020creating}. Dynamics driven by two-body correlations under an effective non-Hermitian Hubbard-like model can also be realized on acoustic platforms, offering an additional avenue for emulation of non-Hermitian physics~\cite{Pu2026}.

As a final particular example, our formalism also applies to arrays of optomechanical sensors, which have been proposed as promising platforms to search for novel and beyond-the-standard-model physics like dark matter~\cite{carney2020proposal,afek2022coherent,brady2023entanglement}. Many-body effects in such systems are a growing focus in optomechanics~\cite{burns1989optical,mohanty2004optical,arita2018optical,vijayan2024cavity}. Recent theoretical and experimental evidence suggests that optical binding forces can include a non-conservative component, opening up a rich vein of physics with fully tunable non-Hermitian, non-reciprocal interactions, parity-time symmetry breaking, and non-equilibrium dynamics~\cite{rudolph2023quantum,livska2023cold,rieser2022tunable,reisenbauer2023non}. Along these lines, for example, Ref.~\cite{xia2023entanglement} proposes using optomechanical sensor arrays that are isolated from one another yet connected via distributed entanglement.


\begin{acknowledgments}R.M.~acknowledges support from the T$_c$SUH Welch Professorship Award. N.A. and J.H.~are 
supported by the Noyce Foundation. R.T.S.~was supported by the grant DOE DE-SC0014671 funded by the U.S.~Department of Energy, Office of Science. Numerical simulations were partially performed with resources provided by the Research Computing Data Core at the University of Houston. The data that support the findings of this article are openly available~\cite{zenodo}.
\end{acknowledgments}


\appendix
\renewcommand{\thefigure}{A\arabic{figure}}
\setcounter{figure}{0}
\renewcommand{\theequation}{A\arabic{equation}}
\setcounter{equation}{0}

\begin{appendix}
\section{Parity-time symmetry} \label{app:pt_symm}

The ladder Hamiltonian considered in the main text is invariant under the combined action of parity $\mathcal P$ and time reversal $\mathcal T$. Here, $\mathcal P$ is defined as spatial inversion about the center of the ladder together with exchange of the two legs, while $\mathcal T$ acts by complex conjugation. Explicitly,
\begin{align}
    (\mathcal{PT})\,\hat c^\dagger_{i,A,\sigma}\,(\mathcal{PT})^{-1}
    &=
    \hat c^\dagger_{L-i,B,\sigma},
    \notag\\
    (\mathcal{PT})\,\hat c^\dagger_{i,B,\sigma}\,(\mathcal{PT})^{-1}
    &=
    \hat c^\dagger_{L-i,A,\sigma}\ ,
    \label{eq:PTops}
\end{align}
and similarly for the annihilation operators,
    \begin{align}
    (\mathcal{PT})\,\hat c_{i,A,\sigma}^{\phantom{\dagger}}\,(\mathcal{PT})^{-1}
    &=
    \hat c_{L-i,B,\sigma}^{\phantom{\dagger}},
    \notag\\
    (\mathcal{PT})\,\hat c_{i,B,\sigma}^{\phantom{\dagger}}\,(\mathcal{PT})^{-1}
    &=
    \hat c_{L-i,A,\sigma}^{\phantom{\dagger}}\ .
\end{align}

Using these relations, the intraleg hopping terms transform as
\begin{align}
    (\mathcal{PT})
    \left(
    \sum_i \hat c^\dagger_{i+1,\alpha,\sigma}\hat c_{i,\alpha,\sigma}^{\phantom{\dagger}}
    \right)
    (\mathcal{PT})^{-1}
    &=
    \sum_i \hat c^\dagger_{i,\bar\alpha,\sigma}\hat c_{i+1,\bar\alpha,\sigma}^{\phantom{\dagger}}\ ,
    \notag\\
    (\mathcal{PT})
    \left(
    \sum_i \hat c^\dagger_{i,\alpha,\sigma}\hat c_{i+1,\alpha,\sigma}^{\phantom{\dagger}}
    \right)
    (\mathcal{PT})^{-1}
    &=
    \sum_i \hat c^\dagger_{i+1,\bar\alpha,\sigma}\hat c_{i,\bar\alpha,\sigma}^{\phantom{\dagger}}\ ,
    \label{eq:PThopping}
\end{align}
where $\bar A=B$ and $\bar B=A$. Likewise, the rung-hopping term transforms as
\begin{equation}
    (\mathcal{PT})
    \left(
    \sum_i \hat c^\dagger_{i,\alpha,\sigma}\hat c_{i,\bar\alpha,\sigma}^{\phantom{\dagger}}
    \right)
    (\mathcal{PT})^{-1}
    =
    \sum_i \hat c^\dagger_{i,\bar\alpha,\sigma}\hat c_{i,\alpha,\sigma}^{\phantom{\dagger}}\ .
    \label{eq:PTrung}
\end{equation}

It then follows that the Hamiltonian is invariant under $\mathcal{PT}$ when the hopping asymmetry is reversed between the two legs, $\delta_A=-\delta_B$, and the remaining couplings, in particular the rung hopping $V_0$ and the onsite interaction $U$, are real, as assumed in the main text. Indeed, under the above transformation, a forward hopping term on one leg is mapped to a backward hopping term on the opposite leg, with exactly the same coefficient appearing in the original Hamiltonian. The rung hopping $V_0$ and the onsite Hubbard interaction are also unchanged. Therefore, 
\begin{equation}
    (\mathcal{PT})\,\hat{\mathcal H}\,(\mathcal{PT})^{-1}
    =
    \hat{\mathcal H},
\end{equation}
and the ladder model is $\mathcal{PT}$ symmetric. This should be contrasted with the single-chain Hatano-Nelson model~\cite{Zhang2022}, for which spatial inversion does not exchange distinct legs and one instead finds $(\mathcal{PT})\hat{\mathcal H}(\mathcal{PT})^{-1}=\hat{\mathcal H}^\dagger$.

A direct consequence of $\mathcal{PT}$ symmetry is that the eigenvalues of $\hat{\mathcal H}$ are either real or occur in complex-conjugate pairs. Indeed, if
\begin{equation}
    \hat{\mathcal H}|\psi\rangle=E|\psi\rangle,
\end{equation}
then, using the antiunitary character of $\mathcal{PT}$ together with $[\hat{\mathcal H},\mathcal{PT}]=0$, one finds
\begin{equation}
    \hat{\mathcal H} \, (\mathcal{PT}|\psi\rangle) = E^* (\mathcal{PT}|\psi\rangle)\ .
\end{equation}
Therefore, $\mathcal{PT}|\psi\rangle$ is also an eigenstate of $\hat{\mathcal H}$, with eigenvalue $E^*$. It follows that a complex eigenvalue must be accompanied by its complex conjugate. A given eigenvalue is
real when the corresponding eigenstate is simultaneously an eigenstate of $\mathcal{PT}$, namely when $\mathcal{PT}|\psi\rangle=|\psi\rangle$ in which case $E=E^*$. This is the unbroken $\mathcal{PT}$-symmetric phase. 
By contrast, when $|\psi\rangle$ is not an eigenstate of $\mathcal{PT}$, the symmetry is said to be broken, and the corresponding eigenvalues form a complex-conjugate pair.

\section{Review of the role of boundary conditions in the reality of the spectrum} \label{app:BC_spec}

In the main text, we discuss the conditions under which the spectrum remains real in the non-interacting regime with periodic boundary conditions (PBC). In non-Hermitian systems, however, the choice of boundary conditions can have a much stronger effect than in Hermitian ones. This is already evident, for example, in the single-chain Hatano-Nelson model~\cite{hatano1996localization,hatano1997vortex},
\begin{equation}
\hat{\mathcal H}_{\rm HN}
=
-\sum_{j,\sigma}
\Big[
(t+\delta)\hat c^\dagger_{j+1,\sigma}\hat c_{j,\sigma}^{\phantom{\dagger}}
+
(t-\delta)\hat c^\dagger_{j,\sigma}\hat c_{j+1,\sigma}^{\phantom{\dagger}}
\Big]\ .
\label{eq:HNsingle}
\end{equation}
Under PBC, its spectrum is generally complex and forms an ellipse in the complex-energy plane. By contrast, under open boundary conditions (OBC), the model can be mapped to a Hermitian tight-binding chain by means of the non-unitary (similarity) transformation
\begin{align}
\hat c_{j,\sigma} &= e^{gj}\,\hat d_{j,\sigma}\ , \notag\\
\hat c^\dagger_{j,\sigma} &= e^{-gj}\,\hat d^\dagger_{j,\sigma}\ , \notag\\
e^{2g} &= \frac{t+\delta}{t-\delta}\ .
\label{eq:OBCgauge}
\end{align}
With this choice, the asymmetric hopping amplitudes are transformed into a symmetric one with effective value
\begin{equation}
\tilde t=\sqrt{(t+\delta)(t-\delta)}=\sqrt{t^2-\delta^2}.
\end{equation}
The Hamiltonian then becomes
\begin{equation}
\hat{\tilde{\mathcal H}}_{\rm HN}
=
-\tilde t\sum_{j,\sigma}
\left(
\hat d^\dagger_{j+1,\sigma}\hat d_{j,\sigma}^{\phantom{\dagger}}
+
\hat d^\dagger_{j,\sigma}\hat d_{j+1,\sigma}^{\phantom{\dagger}}
\right),
\label{eq:HNsingleOBC}
\end{equation}
which is Hermitian.

As a result, the spectrum of the single-chain Hatano-Nelson model is entirely real under OBC as long as $\delta < t$. 
(In the parameterization $te^{\pm h}$ the OBC Hatano-Nelson model has a real eigenspectrum for all real-valued $t,h$.)

If one includes Hubbard-like onsite interactions, i.e., 
\begin{equation}
    \hat{\cal H}_{\rm HN-int} = \hat {\cal H}_{\rm HN} +U\sum_j \hat n_{j,\uparrow} \hat n_{j,\downarrow}\ ,
\end{equation}
where $ \hat n_{j,\sigma}=\hat c^\dagger_{j,\sigma}\hat c_{j,\sigma}^{\phantom{\dagger}}$, the eigenspectrum remains real. This is because the interaction term is purely onsite, as such, the exponential factors cancel exactly, thus its form is unchanged by the transformation and $\tilde{\hat{\cal H}}_{\rm HN-int}$ is Hermitian. Therefore, for the single-chain Hatano-Nelson model, an onsite Hubbard interaction does not affect the reality of the spectrum under OBC.
\begin{figure}[t!]
    \includegraphics[width=1\columnwidth]{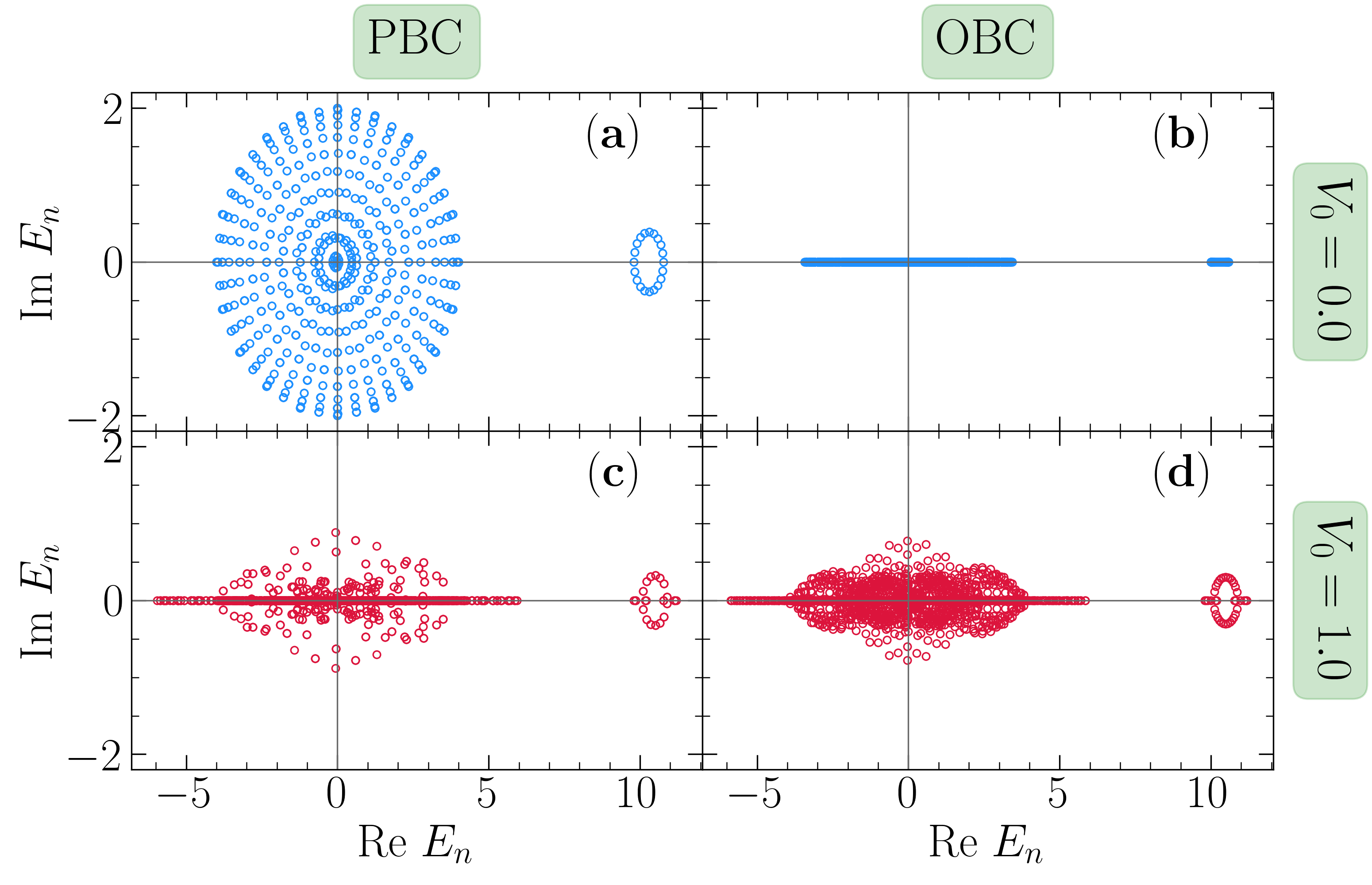}
    \caption{
    Spectrum of $\hat{\mathcal H}$ under two choices of boundary conditions, PBC and OBC, and zero or a finite value of the rung hybridization $V_0$ as indicated. Only when open boundaries are assumed and the hybridization is zero can one obtain a real spectrum, since a similarity transformation to a Hermitian Hamiltonian is possible (see text). Parameters are $\delta=0.5$, $U=10$, and $L=20$, in the filling sector $N_\uparrow=N_\downarrow=1$.}
    \label{fig:BC_comparison}
\end{figure}
This argument does not extend to the two-leg ladder studied in the main text. In that case, one may attempt to symmetrize the nonreciprocal hopping on each leg by using opposite similarity factors on the two chains, consistent with $\delta_A=-\delta_B$. 
That is,
\begin{align}
\hat c_{j,A\sigma} &= e^{gj}\,\hat d_{j,A\sigma}, 
&\qquad
\hat c^\dagger_{j,A\sigma} &= e^{-gj}\,\hat d^\dagger_{j,A\sigma}, \notag\\
\hat c_{j,B\sigma} &= e^{-gj}\,\hat d_{j,B\sigma}, 
&\qquad
\hat c^\dagger_{j,B\sigma} &= e^{gj}\,\hat d^\dagger_{j,B\sigma},
\label{eq:ladder_similarity}
\end{align}
with
\begin{equation}
e^{2g}=\frac{t+\delta}{t-\delta}.
\end{equation}
For $\delta_A=+\delta$ and $\delta_B=-\delta$, this choice symmetrizes the {\it intrachain} hopping on both legs, yielding
\begin{align}
\hat{\tilde{\mathcal K}}_{\parallel}
=
-\tilde t \sum_{j,\sigma}
\Big(
\hat d^\dagger_{j+1,A\sigma}\hat d_{j,A\sigma}
+\hat d^\dagger_{j,A\sigma}\hat d_{j+1,A\sigma}
\notag\\
+\hat d^\dagger_{j+1,B\sigma}\hat d_{j,B\sigma}
+\hat d^\dagger_{j,B\sigma}\hat d_{j+1,B\sigma}
\Big),
\end{align}
where again $\tilde t=\sqrt{t^2-\delta^2}$. The onsite Hubbard term also remains unchanged, since the exponential factors still cancel on each site. In contrast, the rung-hopping term transforms as
\begin{align}
&-V_0\sum_{j,\sigma}
\left(
\hat c^\dagger_{j,A\sigma}\hat c_{j,B\sigma}^{\phantom{\dagger}}
+\hat c^\dagger_{j,B\sigma}\hat c_{j,A\sigma}^{\phantom{\dagger}}
\right)
\nonumber\\
&\quad
=
-V_0\sum_{j,\sigma}
\Big(
e^{-2gj}\hat d^\dagger_{j,A\sigma}\hat d_{j,B\sigma}^{\phantom{\dagger}}
+e^{2gj}\hat d^\dagger_{j,B\sigma}\hat d_{j,A\sigma}^{\phantom{\dagger}}
\Big).
\label{eq:rung_transformed}
\end{align}
Thus, although the nonreciprocal hopping along each individual leg can be symmetrized, the interleg hybridization acquires position-dependent exponential factors. The transformed ladder Hamiltonian is therefore not mapped to a uniform Hermitian ladder, and the single-chain argument for the reality of the spectrum under OBC no longer applies when $V_0\neq 0$.

An example of this is given in Fig.~\ref{fig:BC_comparison} where we compare the eigenspectrum of $\hat {\cal H}$ for the two different boundary conditions and for finite and vanishing interleg (rung) hybridization at a finite interaction $U =10$ for $N_\uparrow=N_\downarrow=1$. As indicated above, one generically expects a real spectrum only when dealing with OBC and for $V_0=0$.

\section{Finite-size effects} \label{app:FSE}
\begin{figure}[t!]
\includegraphics[width=\columnwidth]{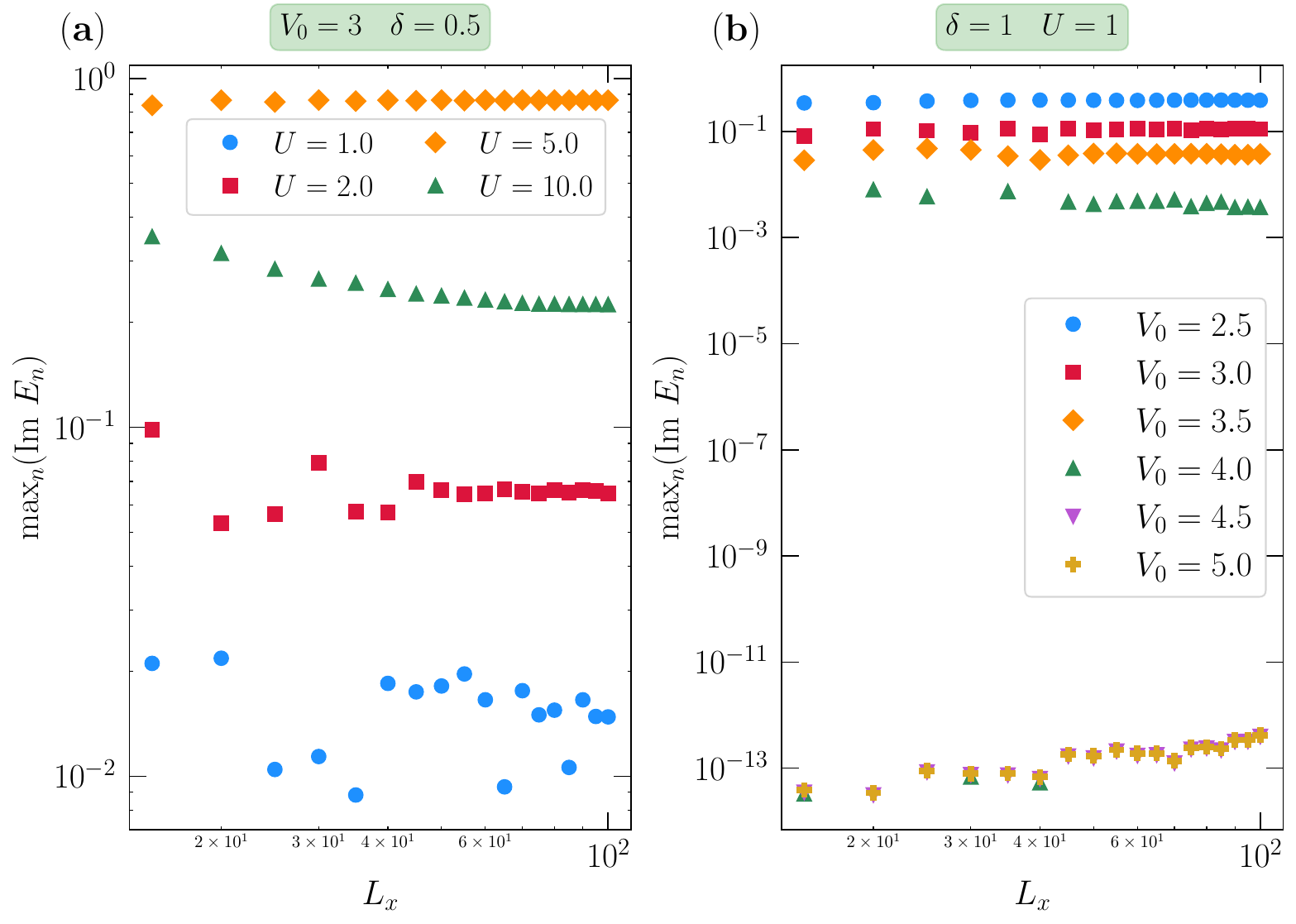}
    \caption{(a) Finite-size scaling of the largest imaginary part of the spectrum,     $\max_n ({\rm Im}\ E_n)$, for $V_0 = 3$ and $\delta = 0.5$ at different interaction strengths. For $U = 1$, the values fluctuate around $10^{-2}$, showing weak size dependence, while for $U\ge 2$, the imaginary parts saturate to size-independent constants, indicating that the complex spectrum persists in the thermodynamic limit. (b) The same for $\delta = 1$, $U=1$ and $V_0$ ranging from 2.5 to 5.
    }
    \label{fig:fig_fse}
\end{figure}

To check that our results on whether the eigenspectrum $\{E_n\}$ exhibits imaginary values or not are representative of the thermodynamic limit, we took fixed $V_0$ [Fig.~\ref{fig:fig_fse}(a)] and fixed $U$ [Fig.~\ref{fig:fig_fse}(b)] one-dimensional cuts of the ``phase" diagram and performed finite-size scaling analysis. Fig.~\ref{fig:fig_fse}(a) shows the $\max_n({\rm Im} \ E_n)$ for various interaction strengths, as a vertical cut in the original Fig.~\ref{fig:fig_2}(c) at $V_0=3$. Apart from small fluctuations at $U=1$, there is clear convergence with increasing system size, including the non-monotonicity with $U$. In turn, Fig.~\ref{fig:fig_fse}(b) corresponds to a horizontal cut in Fig.~\ref{fig:fig_2}(a) at fixed $\delta =1$. Here, it is clear that the transition point, at around $V_0 \simeq 4$ and stemming from the continua of two-particle bands, survives in the thermodynamic limit.

\section{Two-particle continua in the Hermitian limit \texorpdfstring{$\delta=0$}{delta=0}} \label{app:two_non_int_particles}

It is useful to understand the origin of the characteristic scale $V_0=4t$ appearing in the weakly interacting regime from the Hermitian limit $\delta=0$ [see Fig.~\ref{fig:fig_2}(a)]. In this case, the single-particle Bloch Hamiltonian is 
\begin{equation}
    H_k=
    \begin{pmatrix}
    -2t\cos k & -V_0 \\
    -V_0 & -2t\cos k
    \end{pmatrix},
\end{equation}
whose eigenvalues are $\varepsilon_\pm(k)=-2t\cos k \pm V_0$. These bands correspond to the bonding ($-$) and antibonding ($+$) combinations of the two legs.

In the noninteracting two-particle sector, the eigenstates are products of single-particle states, so the corresponding energies are simply sums of one-particle band energies, $E_{\mu\nu}(k_1,k_2)=\varepsilon_\mu(k_1)+\varepsilon_\nu(k_2)$ with $\mu,\nu\in\{+,-\}$. Introducing the total and relative momenta,
\begin{equation}
    K=k_1+k_2,
    \qquad
    q=\frac{k_1-k_2}{2},
\end{equation}
so that
\begin{equation}
    k_1=\frac{K}{2}+q,
    \qquad
    k_2=\frac{K}{2}-q,
\end{equation}
and using
\begin{equation}
\cos\!\left(\frac{K}{2}+q\right)+\cos\!\left(\frac{K}{2}-q\right)
=
2\cos\!\left(\frac{K}{2}\right)\cos q,
\end{equation}
one obtains three distinct bands,
\begin{align}
E_{--}(K,q)
&=
-4t\cos\!\left(\frac{K}{2}\right)\cos q -2V_0, 
\label{eq:Eminusminus_appendix} 
\\
E_{+-}(K,q)
&=
-4t\cos\!\left(\frac{K}{2}\right)\cos q,
\label{eq:Eplusminus_appendix} \\
E_{++}(K,q)
&=
-4t\cos\!\left(\frac{K}{2}\right)\cos q +2V_0. 
\label{eq:Eplusplus_appendix}
\end{align}
The mixed sectors $(+,-)$ and $(-,+)$ give the same set of energies and are
therefore represented by the single branch $E_{+-}(K,q)$.

Each band has a width $8t$, since
\begin{equation}
-1 \le \cos\!\left(\frac{K}{2}\right)\cos q \le 1,
\end{equation}
and is centered at $-2V_0$, $0$, and $+2V_0$, respectively. Hence their energy
supports are
\begin{align}
E_{--} &\in [-4t-2V_0,\; 4t-2V_0], \notag
\\
E_{+-} &\in [-4t,\; 4t],
\\
E_{++} &\in [-4t+2V_0,\; 4t+2V_0]. \notag
\end{align}
A direct consequence is that the three bands open their gaps only when $V_0>4t$. This provides a natural explanation for the characteristic scale $V_0/t\simeq 4$ that appears in the weakly interacting regime of the main text: for $V_0<4t$, the Hermitian two-particle continua overlap, so even a small non-Hermitian perturbation can mix them and generate complex-conjugate eigenvalue pairs, whereas for $V_0>4t$, a finite non-Hermitian strength is
required before the spectrum acquires an imaginary part.

\section{Same and reversed fluxes in the winding-number calculation} \label{app:same_opp_flux}

\begin{figure}[t!]
    \includegraphics[width=1\columnwidth]{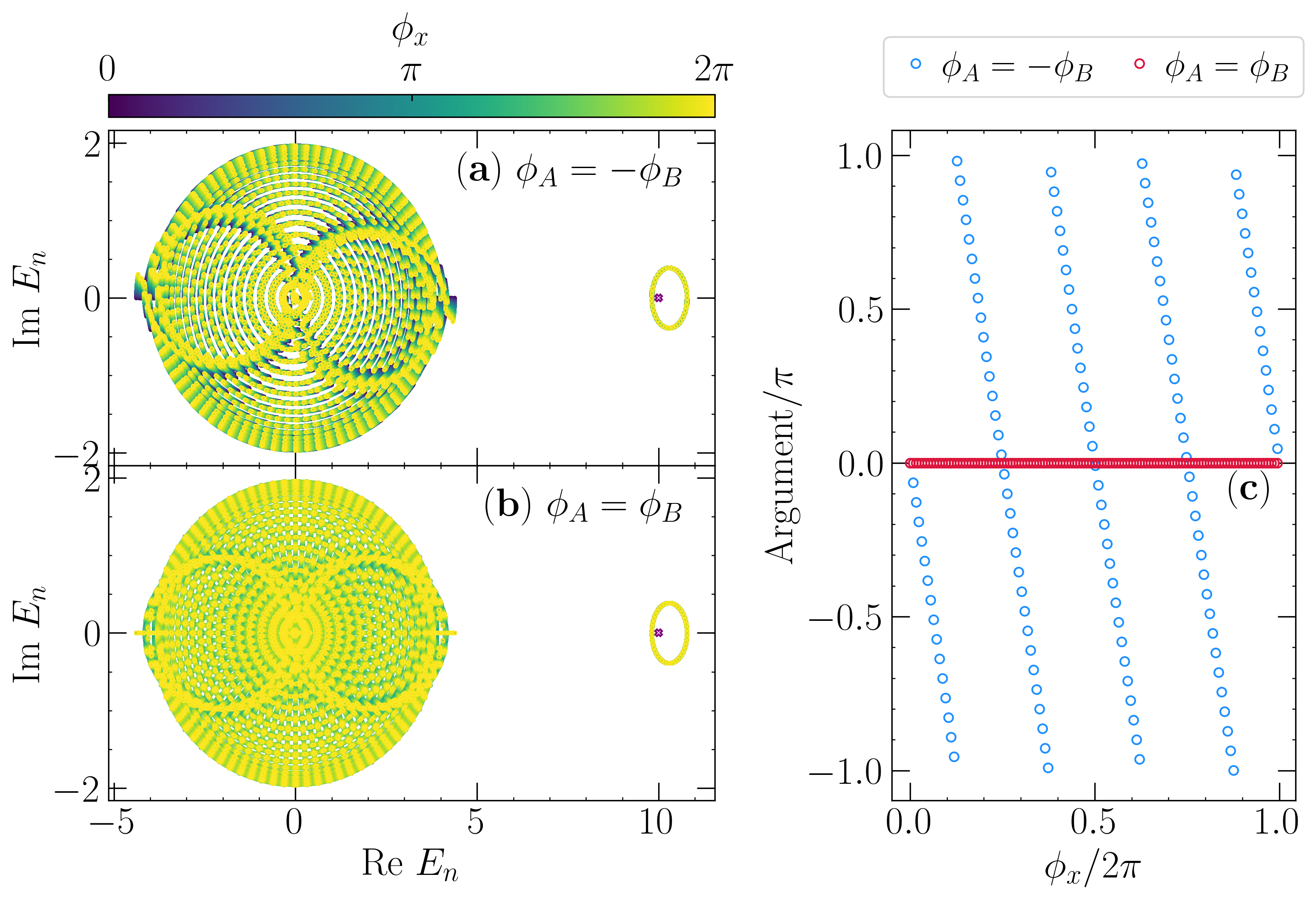}
    \caption{
    Spectrum of $\hat{\mathcal H}(\phi)$ under two choices of probing flux. In (a), the fluxes are opposite on the two legs, $\phi_A=-\phi_B$, so that the net flux through the ladder is zero. In (b), the fluxes are identical, $\phi_A=\phi_B$. The color scale indicates the magnitude of the applied flux. Panel (c) shows the corresponding argument of $\det[\hat{\mathcal H}(\phi)-E]$ as a function of the flux, for the target energy $E=U$ marked in panels (a) and (b). Only the opposite-flux choice yields the net phase winding corresponding to $W=4$. Parameters are $\delta=0.5$, $V_0=0.2$, $U=10$, and $L=40$, in the sector $N_\uparrow=N_\downarrow=1$.
    }
    \label{fig:same_opp_flux}
\end{figure}

In the main text, we argued that the spectral winding number should be probed by a flux pattern consistent with the asymmetry of the non-Hermitian hopping. For the ladder studied here, where the hopping nonreciprocity is opposite on the two legs, $\delta_A=-\delta_B$, this corresponds to choosing opposite probing fluxes,
$\phi_A=-\phi_B$, rather than identical ones, $\phi_A=\phi_B$, in Eq.~\eqref{eq:Kphi}. The two choices are compared in Fig.~\ref{fig:same_opp_flux}.

As shown there, only the opposite-flux choice captures the finite total winding of the high-energy spectral ring, centered around $E\simeq U$. For the parameters of Fig.~\ref{fig:same_opp_flux}, this point gap has winding number $W=4$. By contrast, choosing the same flux in both legs does not reproduce this winding structure. Thus, in the present model, the probing flux must follow the pattern of reversed nonreciprocity, $\delta_A=-\delta_B$, in order to diagnose the relevant point-gap topology.

\section{Winding number at higher densities}
\label{app:w_at_higher_densities}

\begin{figure}[t!]
    \includegraphics[width=1\columnwidth]{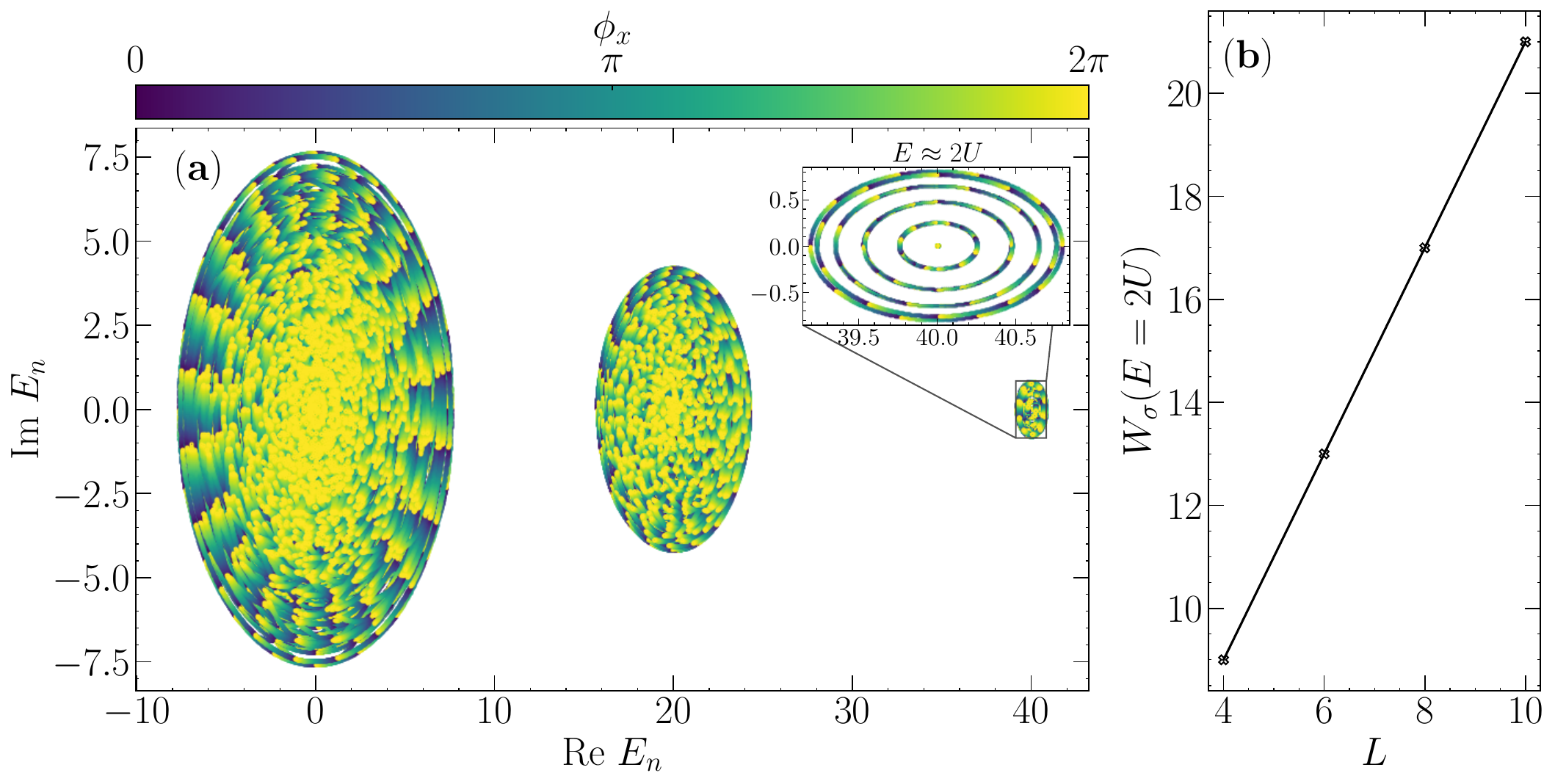}
    \caption{(a) Spectrum of the Hamiltonian with flux inserted only in the spin-$\uparrow$ sector, for $N_\uparrow=N_\downarrow=2$; the inset shows a zoom around $E\approx 2U$. Parameters are $L=10$, $V_0=0.1$, $\delta=1$, and $U=20$. (b) Spin-resolved winding number at $E=2U$ as a function of $L$, showing that, unlike the $N_\uparrow=N_\downarrow=1$ case at $E=U$ in the main text, this quantity is no longer size-independent.
    }
    \label{fig:higher_filling_winding}
\end{figure}

In the main text, we focused on the winding number in the two-particle sector $N_\uparrow=N_\downarrow=1$. Here we briefly contrast that case with slightly higher fillings. We begin with $N_\uparrow=N_\downarrow=2$ [Fig.~\ref{fig:higher_filling_winding}], where at strong interactions ($U=20$) the spectrum separates into three main structures: a low-energy sector with predominantly no doublons, a second structure centered near $E\approx U$ with predominantly single-doublon character, and a third one near $E\approx 2U$ associated with states containing two doublons. In this last case, unlike the $N_\uparrow=N_\downarrow=1$ sector discussed in the main text, the spin-resolved winding number at $E=2U$ is no longer size-independent, but instead grows with
system size [Fig.~\ref{fig:higher_filling_winding}(b)].

\begin{figure}[b!]
    \includegraphics[width=0.95\columnwidth]{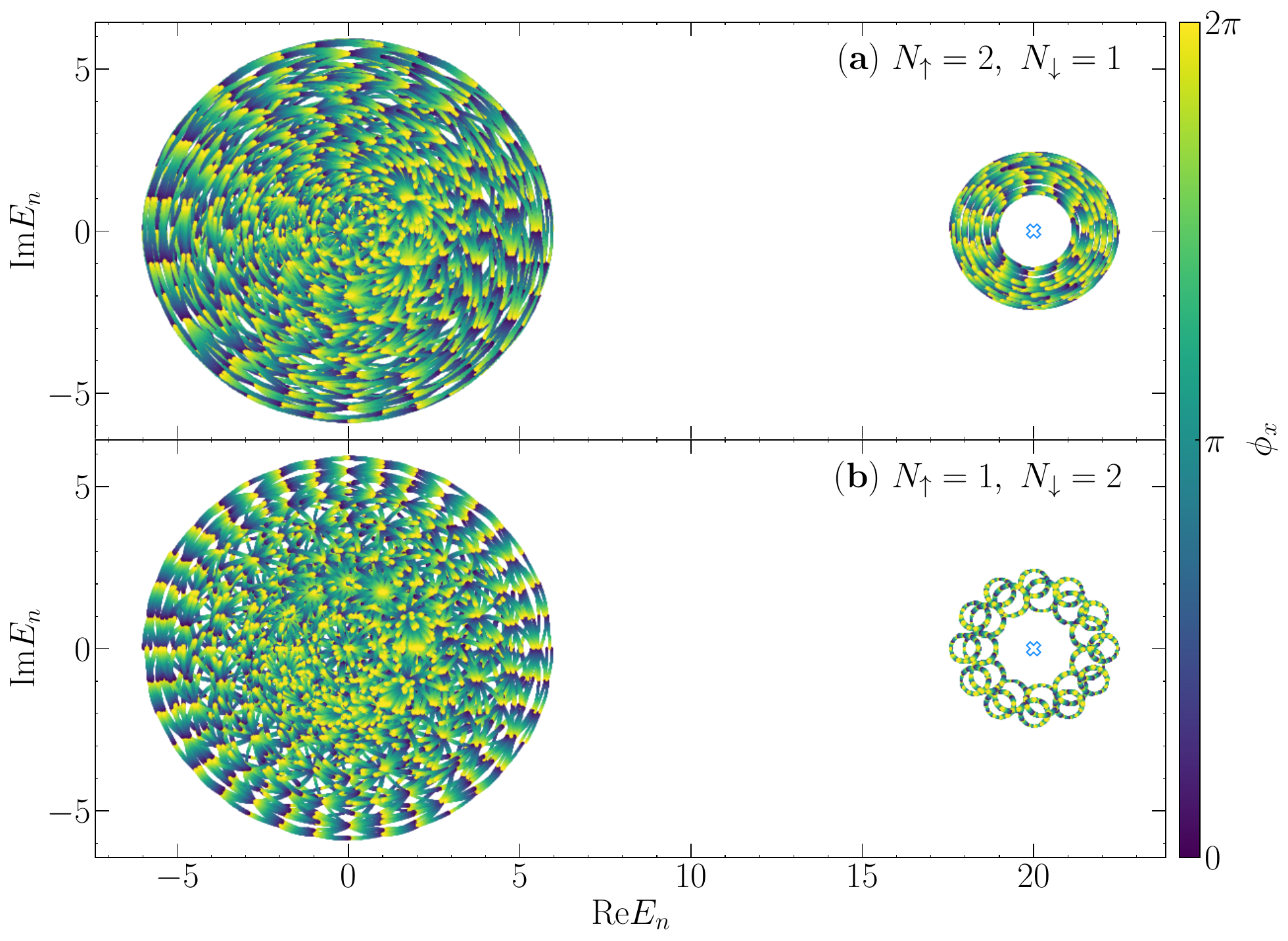}
    \caption{Flux-resolved spectra with flux inserted only in the spin-$\uparrow$ sector for (a) $N_\uparrow=2$, $N_\downarrow=1$ and (b) $N_\uparrow=1$, $N_\downarrow=2$. Parameters are $L=12$, $V_0=0.1$,     $\delta=1$, and $U=20$. The marker at $E=U$ indicates the target energy of the one-doublon sector.
    }
    \label{fig:unbalanced_filling_winding}
\end{figure}

A less dramatic modification is to add a single particle to the $N_\uparrow=N_\downarrow=1$ sector, leading to the imbalanced fillings $N_\uparrow=2$, $N_\downarrow=1$ and $N_\uparrow=1$, $N_\downarrow=2$. In these cases, the spectrum reverts to two main structures, corresponding roughly to zero- and one-doublon sectors; see Fig.~\ref{fig:unbalanced_filling_winding}. For $N_\uparrow=2$, $N_\downarrow=1$, the flux-resolved spectrum of $\hat{\mathcal H}_\uparrow$ still exhibits a detached structure near $E\approx U$. By contrast, for $N_\uparrow=1$, $N_\downarrow=2$, the fluxed spin-$\uparrow$ spectrum does not contain an isolated loop enclosing $E=U$, and the corresponding winding therefore vanishes, $W_\uparrow(E=U)=0$. We have also checked that the spin-resolved winding in the $N_\uparrow=2$, $N_\downarrow=1$ case is not size-independent (not shown).

\section{Collapse of the winding number at large interleg hybridization}
\label{app:large_V0}

\begin{figure}[h!]
    \includegraphics[width=1\columnwidth]{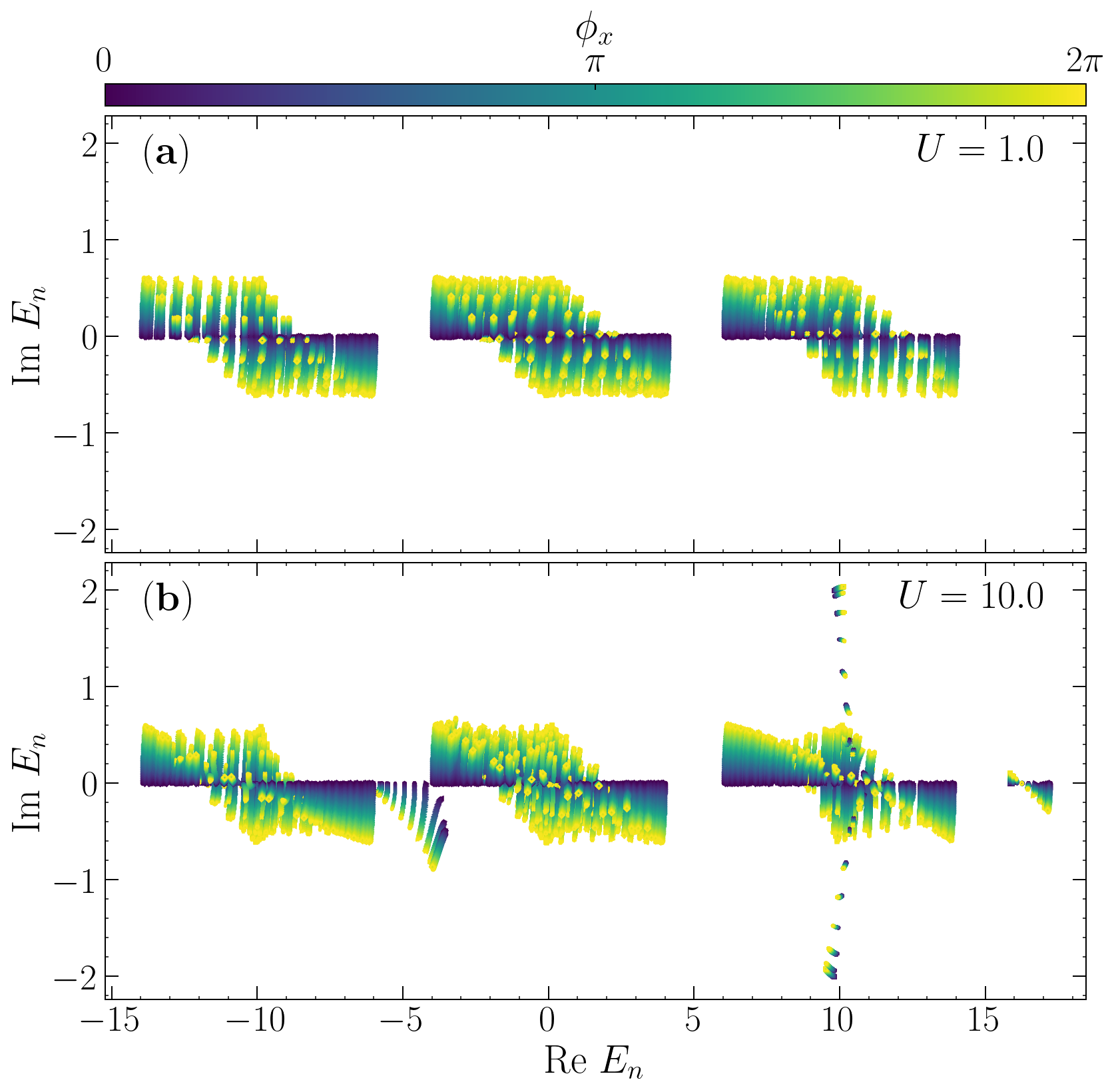}
    \caption{Flux-resolved spectra in the $N_\uparrow=N_\downarrow=1$ sector at fixed $\delta=1$ and $V_0=5$ for an $L=20$ ladder, showing the loss of the simple isolated point-gap structure present at weak interleg hybridization. (a) $U=1$: the spectrum is organized into broad spectral ribbons, with no detached high-energy loop around $E\simeq U$. (b) $U=10$: a high-energy structure near ${\rm Re}\,E_n\sim U$ remains visible, but it is no longer represented by a single isolated loop and coexists with additional high-energy features. In both cases, the flux-resolved spectrum no longer supports the simple doublon winding discussed in the main text.
}
    \label{fig:large_V0}
\end{figure}

Figure~\ref{fig:fig_4} in the main text illustrates the flux-resolved spectrum for weak interleg hybridization, $V_0=0.1$, where the detached doublon branch encloses a clear point gap and therefore carries a well-defined winding number. At larger interleg hybridization, however, this simple structure is lost. Representative examples are shown in Fig.~\ref{fig:large_V0} for $V_0=5$ and $\delta=1$.

For small interactions, $U=1$ [Fig.~\ref{fig:large_V0}(a)], the flux-resolved spectrum is organized into broad spectral ribbons and no longer contains a detached high-energy loop around $E\simeq U$. In this case, the point-gap structure used to define the doublon winding in the weak-$V_0$ regime is absent.

For stronger interactions, $U=10$ [Fig.~\ref{fig:large_V0}(b)], a high-energy structure near ${\rm Re}\,E_n\sim U$ remains visible, but it is no longer represented by a single isolated loop of the type found at weak $V_0$. Additional high-energy structures appear, and the corresponding point-gap picture becomes less robust. Thus, at large interleg hybridization, the collapse of the winding number should be understood as the loss of the isolated doublon point gap, rather than simply as the disappearance of all imaginary
parts of the spectrum, which only takes place for a flux $\phi=0$.

\section{Higher Densities}\label{app:Many_Body}

\begin{figure}[t!]
    \includegraphics[height=3.0in,width=3.6in]{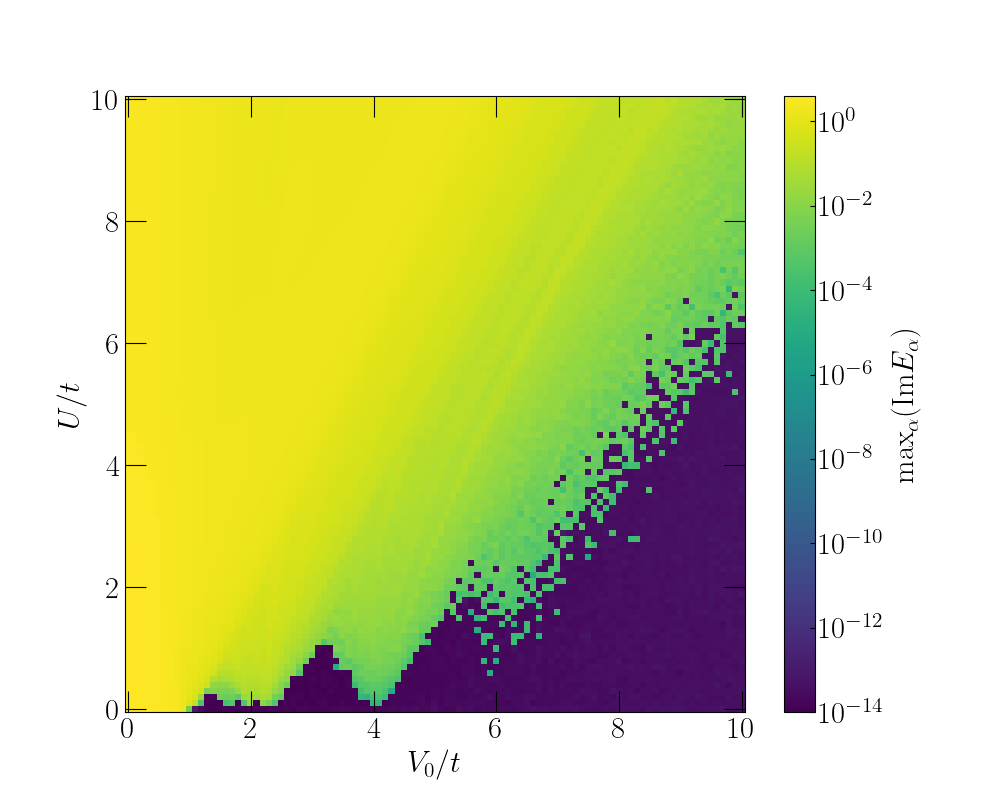}
    \caption{Largest imaginary part of all eigenvalues over all particle sectors
    at fixed $\delta=0.5$, $L=4$ and varying $U, V$.}
    \label{fig:S1}
\end{figure}

At half-filling $N_\uparrow=N_\downarrow=N/2$, the size of the Hilbert space of Hubbard models grows exponentially with lattice size $N$. For this reason, the main text focused on the lower density $N_\uparrow=N_\downarrow=1$ sector.  For small $N$, however, we can see whether some of our conclusions still apply at half-filling. Figure  \ref{fig:S1} undertakes that goal.  It is the analog of Fig.~\ref{fig:fig_2}(c)- a heat map of the imaginary part of the eigenvalues in the plane of the Hubbard interaction $U/t$ and interchain hybridization $V_0/t$, but for {\it all eigenvalues}, i.e.~over all particle number sectors of the Hilbert space. The data appear `noisier', as is typical for small lattice studies.  However, the basic structure is similar: a real spectrum is favored at weak interaction $U$, but when $V_0/t \gtrsim 4$, the interchain hybridization increasingly suppresses the imaginary part of the eigenvalues.  Unsurprisingly, this `protection' of the real spectrum extends to smaller values of $U/t$ for the
complete Hilbert space of all particle numbers.

\section{OBC Spectrum}\label{app:OBC}

\begin{figure}[t!]
    \includegraphics[height=3.0in,width=3.6in]{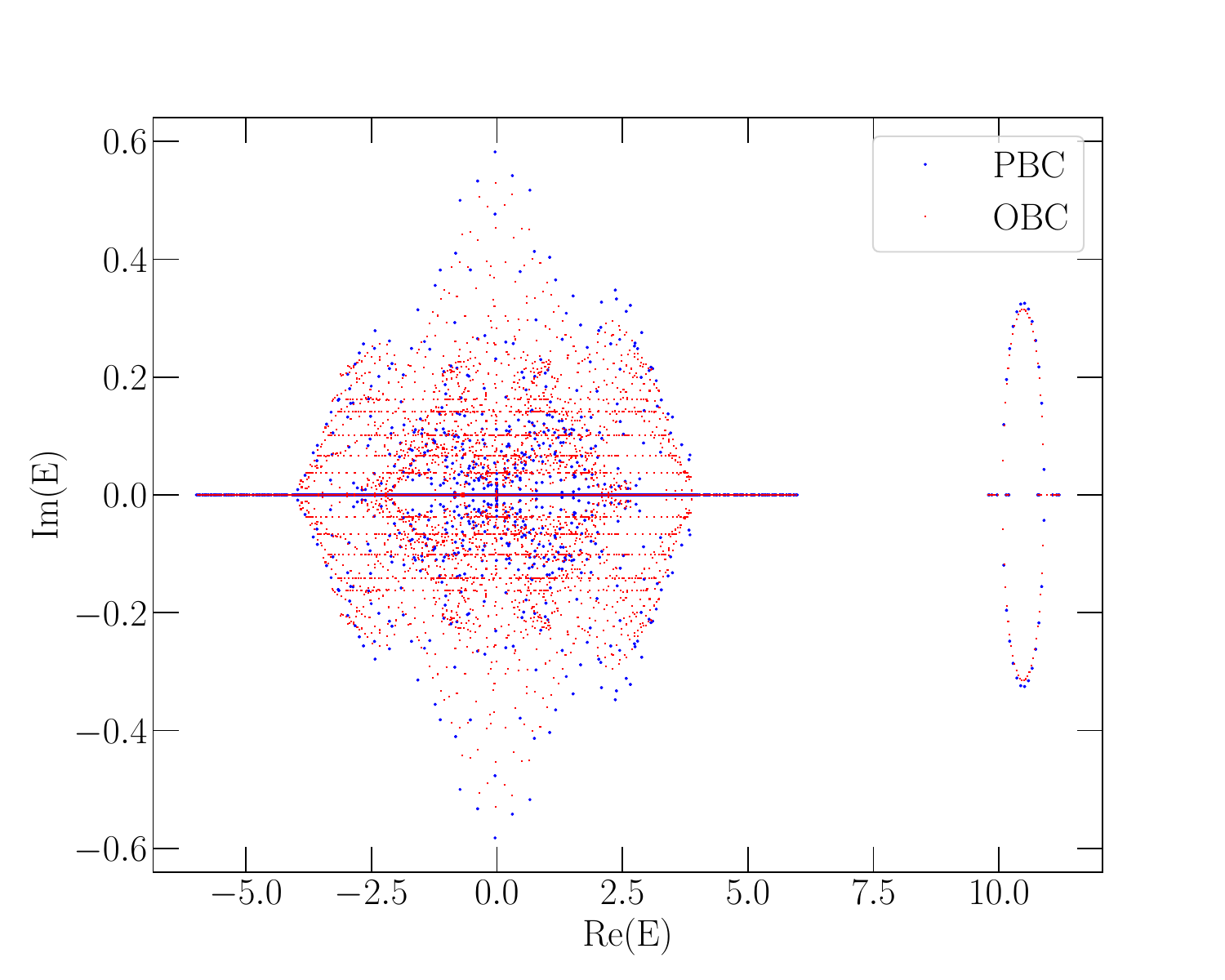}
    \caption{Comparison of PBC and OBC eigenvalues for the $40\times2$ system at the $N_\uparrow=N_\downarrow=1$ sector at $U=10$, $\delta=0.5$, $V_0=1$. 
    }
    \label{fig:S6}
\end{figure}

The eigenspectrum analysis in the main text focuses on the ring cases. To understand how the boundary conditions affect it, Figure \ref{fig:S6} contrasts the PBC and OBC spectra for a particular parameter set, $U=10$, $\delta=0.5$, $V_0=1$ for the $N_\uparrow=N_\downarrow$ filling on a $N=40 \times 2$ lattice.  While the precise eigenvalue locations in the complex plane are shifted (as is expected since the discrete momenta $k_m$ are different), the overall topologies are remarkably similar.  There is a small group of eigenvalues forming a ring about the point $(U,0)$.  These correspond to the $N=80$ basis states with double occupation. Then there is a larger grouping of $N(N-1)=80\times79$ eigenvalues centered at $(0,0)$ arising from the basis states in which the up and down electrons occupy distinct sites. In both cases, some eigenvalues have collapsed onto the real axis. The emergence of real eigenvalues originates from the formation of layer-hybridized states as $V_0$ increases.

\begin{figure}[t!]
    \includegraphics[height=3.0in,width=3.6in]{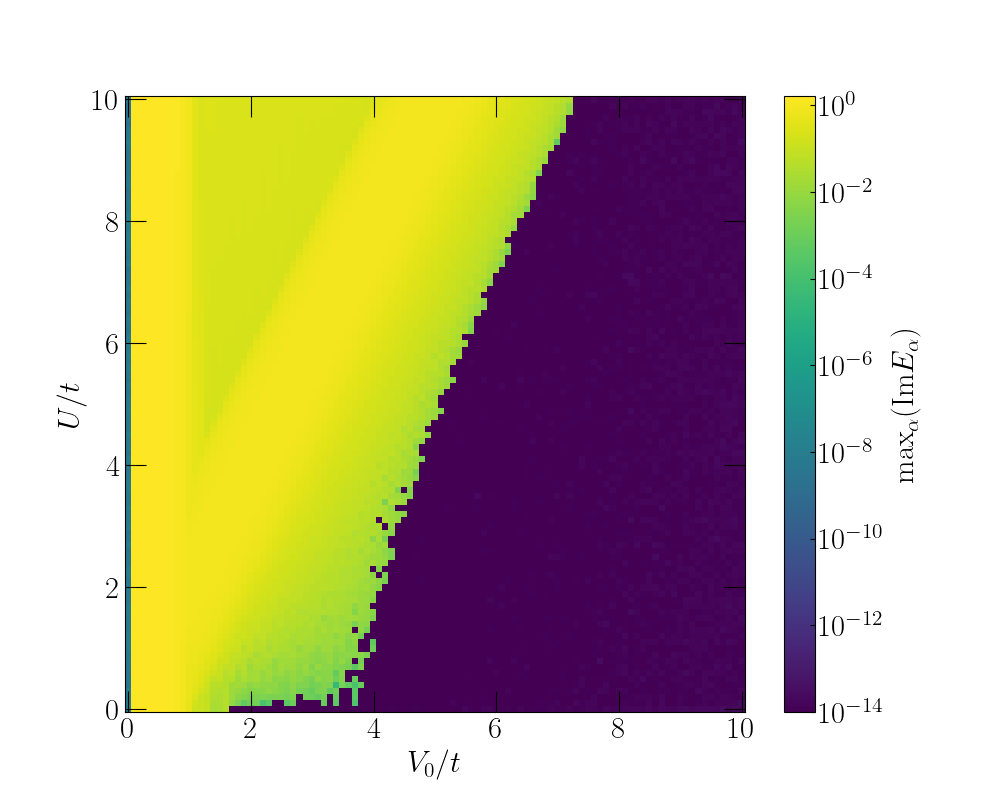}
    \includegraphics[height=3.0in,width=3.6in]{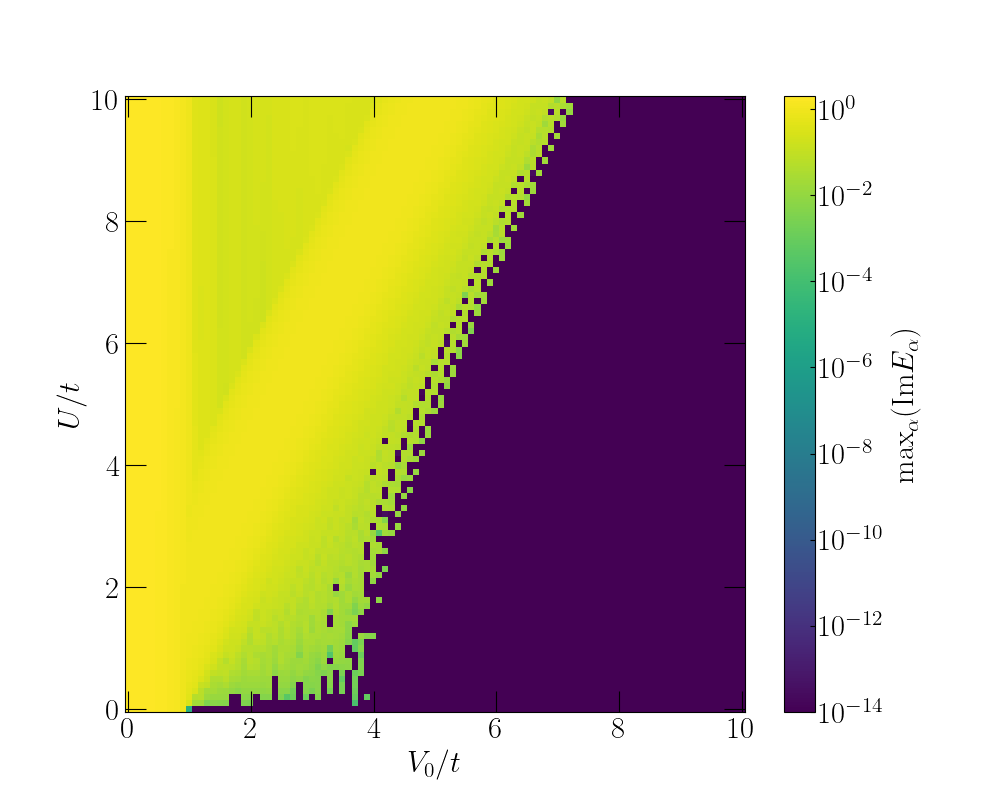}
    \caption{
    Top: Heat map of the largest imaginary part of all eigenvalues for OBC. There are two critical values for the real-complex transition, as one has the trivial line $V_0=0$. For any small finite $V_0$ away from the critical decoupled limit, the spectrum becomes complex. The real spectrum then re-emerges for large $V_0$.
    Bottom: Heat map for the PBC case for comparison. Results are from the $N_\uparrow=N_\downarrow=1$ sector at fixed $\delta=0.5$, $L=20$ in the $V_0-U$ plane.
    }
    \label{fig:S3}
\end{figure}

The sensitivity of the Hatano-Nelson spectrum to boundary conditions is well known in the non-interacting, decoupled one-dimensional case, as reviewed in the main text and in Appendix~\ref{app:BC_spec}.  In contrast, and consistent with Fig.~\ref{fig:S6}, Fig.~\ref{fig:S3} emphasizes that the OBC complex-real ``phase" diagram has the same features as its PBC counterpart, Fig.~\ref{fig:fig_2}(c) in the main text.  Specifically, both plots show a critical point for a real spectrum at $V_0 = 2\delta$ for $U=0$.  When the on-site interaction $U$ is small, but non-zero, regardless of boundary condition, the requirement for a real spectrum jumps to $V_0 \gtrsim 4$. The critical interchain hybridization grows with $U$, reaching, in both cases, $V_0/t \gtrsim 7$
at the maximum repulsion shown, $U/t=10$.
The only substantial difference is along the vertical axis, $V_0=0$, corresponding to two decoupled chains.  There the spectrum is complex for PBC and purely real for OBC.

A simple two-site interacting model in the $N_\uparrow=N_\downarrow=1$ sector already reveals some insight into the trend we observe for OBC. The Hamiltonian matrix takes the form
\begin{equation}
    H=\begin{pmatrix}
        U &t+\delta & -(t+\delta)& 0\\
        t-\delta & 0 & 0 & t-\delta\\
        -(t-\delta) & 0 & 0 & -(t-\delta)\\
        0 & t+\delta & -(t+\delta)& U
    \end{pmatrix}
\end{equation}
whose four eigenvalues are given by 
\begin{equation}
    \lambda=0,U
\end{equation}
and
\begin{equation}
    \lambda^2-U\lambda-4(t^2-\delta^2)=0
\end{equation}
which has entirely real roots if 
\begin{equation}
    \sqrt{1+\frac{(U/t)^2}{16}}\geq(\delta/t)
\end{equation}
assuming $t,U$ and $\delta$ are positive without loss of generality. This two-site calculation shows the effect of the Hubbard $U$ interaction in the OBC case in the decoupled ($V_0=0$) limit. For $\delta<t$, one always obtains a real spectrum for any finite $U$. In fact, increasing $U$ even favors the onset of a real spectrum when one is in the $\delta>t$ regime where the standard similarity ``gauge" transformation in the 1D OBC case (Appendix~\ref{app:BC_spec}) is not available, as one sees that the spectrum remains real regardless.

\section{Many Layer (2D) System} \label{app:Bulk}
\begin{figure}[t!]
    \includegraphics[height=3.0in,width=3.6in]{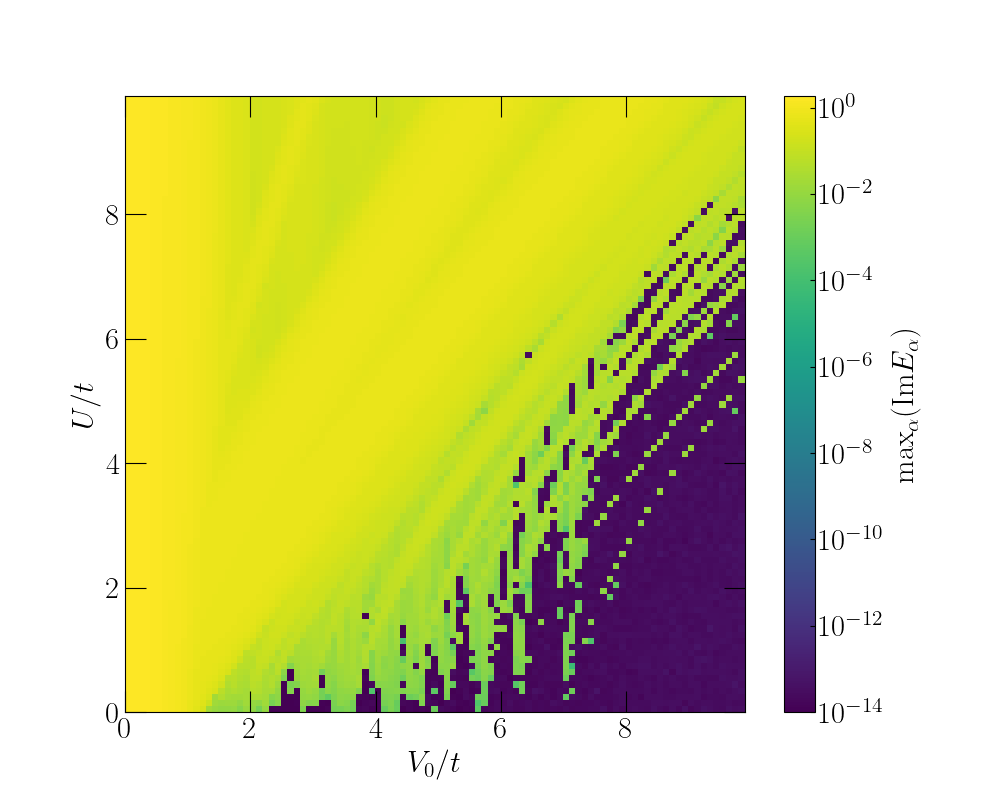}
    \caption{Largest imaginary part of all eigenvalues for the $10\times4$ bulk system at the $N_\uparrow=N_\downarrow=1$ sector, for fixed $\delta=0.5, V=0.5$ and varying $U, V_0$.}
    \label{fig:S5}
\end{figure}

We have also considered a bulk Hamiltonian comprised of two Hermitian 
layers sandwiched between two non-Hermitian layers with opposite
asymmetrical hopping strengths,
\begin{align}
\hat{\mathcal H}_B ={}&
-\sum_{j,\sigma}\sum_{\alpha=A,D}
\Big[
(t+\delta_\alpha)\,
\hat c^\dagger_{j+1,\alpha\sigma}\hat c_{j,\alpha\sigma}
\nonumber\\
&\hspace{1.4cm}
+(t-\delta_\alpha)\,
\hat c^\dagger_{j,\alpha\sigma}\hat c_{j+1,\alpha\sigma}
\Big]
\nonumber\\
&-V_0\sum_{j,\sigma}
\Big(
\hat c^\dagger_{j,A\sigma}\hat c_{j,B\sigma}
+\hat c^\dagger_{j,C\sigma}\hat c_{j,D\sigma}
+\mathrm{h.c.}
\Big)
\nonumber\\
&-V\sum_{j,\sigma}
\Big(
\hat c^\dagger_{j,B\sigma}\hat c_{j,C\sigma}
+\mathrm{h.c.}
\Big)
\nonumber\\
&+U\sum_{j,\alpha}
\hat n_{j,\alpha\uparrow}\hat n_{j,\alpha\downarrow}.
\label{eq:Bulk}
\end{align}
with the layer index $\alpha=A,B,C,D$ ordered from the topmost to bottommost layer. We consider a constant inter-layer coupling $V=0.5$ between the Hermitian layers $B,C$ and vary the coupling $V_0$ to the non-Hermitian edges (from layers $A$ to $B$ and layers $C$ to $D$). We find the same qualitative result in the transition to a completely real spectrum [Fig.~\ref{fig:S5}] at a large enough critical $V_0$. The qualitative robustness of the real-valued ``phase" diagram in the presence of many-body interactions is retained, but a more complex fringe pattern emerges. 

\end{appendix}

\bibliography{references,aggarwal}

@article{xia2023entanglement,
  title={Entanglement-enhanced optomechanical sensing},
  author={Xia, Yi and Agrawal, Aman R and Pluchar, Christian M and Brady, Anthony J and Liu, Zhen and Zhuang, Quntao and Wilson, Dalziel J and Zhang, Zheshen},
  journal={Nature Photonics},
  volume={17},
  number={6},
  pages={470--477},
  year={2023},
  publisher={Nature Publishing Group UK London},
  url={https://www.nature.com/articles/s41566-023-01178-0.pdf}
}

@article{reisenbauer2023non,
  title={Non-{H}ermitian dynamics and nonreciprocity of optically coupled nanoparticles},
  author={Reisenbauer, Manuel and Rudolph, Henning and Egyed, Livia and Hornberger, Klaus and Zasedatelev, Anton V and Abuzarli, Murad and Stickler, Benjamin A and Deli{\'c}, Uro{\v{s}}},
  journal={arXiv preprint arXiv:2310.02610},
  url={https://arxiv.org/pdf/2306.11893},
  year={2023}
}

@article{rudolph2023quantum,
  title={Quantum theory of non-{H}ermitian optical binding between nanoparticles},
  author={Rudolph, Henning and Deli{\'c}, Uro{\v{s}} and Hornberger, Klaus and Stickler, Benjamin A},
  journal={arXiv preprint arXiv:2306.11893},
  year={2023},
  url={https://arxiv.org/pdf/2306.11893}
}

@article{rieser2022tunable,
  title={Tunable light-induced dipole-dipole interaction between optically levitated nanoparticles},
  author={Rieser, Jakob and Ciampini, Mario A and Rudolph, Henning and Kiesel, Nikolai and Hornberger, Klaus and Stickler, Benjamin A and Aspelmeyer, Markus and Deli{\'c}, Uro{\v{s}}},
  journal={Science},
  volume={377},
  number={6609},
  pages={987--990},
  year={2022},
  url={https://www.science.org/doi/pdf/10.1126/science.abp9941},
  publisher={American Association for the Advancement of Science}
}

@article{livska2023cold,
  title={Cold damping of levitated optically coupled nanoparticles},
  author={Li{\v{s}}ka, Vojt{\v{e}}ch and Zem{\'a}nkov{\'a}, Tereza and Svak, Vojt{\v{e}}ch and J{\'a}kl, Petr and Je{\v{z}}ek, Jan and Br{\'a}neck{\`y}, Martin and Simpson, Stephen H and Zem{\'a}nek, Pavel and Brzobohat{\`y}, Oto},
  journal={Optica},
  volume={10},
  number={9},
  pages={1203--1209},
  year={2023},
  url={https://opg.optica.org/directpdfaccess/81ba20f3-b6bf-4a3d-b153d41c90451924_537233/optica-10-9-1203.pdf?da=1&id=537233&seq=0&mobile=no},
  publisher={Optica Publishing Group}
}

@article{burns1989optical,
  title={Optical binding},
  author={Burns, Michael M and Fournier, Jean-Marc and Golovchenko, Jene A},
  journal={Physical Review Letters},
  volume={63},
  number={12},
  pages={1233},
  year={1989},
  url={ttps://journals.aps.org/prl/abstract/10.1103/PhysRevLett.63.1233},
  publisher={APS}
}

@article{mohanty2004optical,
  title={Optical binding between dielectric particles},
  author={Mohanty, Samarendra Kumar and Andrews, Joseph Thomas and Gupta, Pradeep Kumar},
  journal={Optics Express},
  volume={12},
  number={12},
  pages={2746--2753},
  year={2004},
  url={https://opg.optica.org/directpdfaccess/a3df37e6-3fe8-4858-921054a941d20819_80249/oe-12-12-2746.pdf?da=1&id=80249&seq=0&mobile=no},
  publisher={Optica Publishing Group}
}

@article{arita2018optical,
  title={Optical binding of two cooled micro-gyroscopes levitated in vacuum},
  author={Arita, Yoshihiko and Wright, Ewan M and Dholakia, Kishan},
  journal={Optica},
  volume={5},
  number={8},
  pages={910--917},
  year={2018},
  url={https://opg.optica.org/directpdfaccess/c92bbc0f-3cef-4ec2-8f2d19efe39e6bf2_395665/optica-5-8-910.pdf?da=1&id=395665&seq=0&mobile=no},
  publisher={Optica Publishing Group}
}

@article{vijayan2024cavity,
  title={Cavity-mediated long-range interactions in levitated optomechanics},
  author={Vijayan, Jayadev and Piotrowski, Johannes and Gonzalez-Ballestero, Carlos and Weber, Kevin and Romero-Isart, Oriol and Novotny, Lukas},
  journal={Nature Physics},
  volume={20},
  pages={1--6},
  year={2024},
  url={https://www.nature.com/articles/s41567-024-02405-3.pdf},
  publisher={Nature Publishing Group UK London}
}

@article{carney2020proposal,
  title={Proposal for gravitational direct detection of dark matter},
  author={Carney, Daniel and Ghosh, Sohitri and Krnjaic, Gordan and Taylor, Jacob M},
  journal={Physical Review D},
  volume={102},
  number={7},
  pages={072003},
  year={2020},
  url={https://journals.aps.org/prd/pdf/10.1103/PhysRevD.102.072003},
  publisher={APS}
}

@article{brady2023entanglement,
  title={Entanglement-enhanced optomechanical sensor array with application to dark matter searches},
  author={Brady, Anthony J and Chen, Xin and Xia, Yi and Manley, Jack and Dey Chowdhury, Mitul and Xiao, Kewen and Liu, Zhen and Harnik, Roni and Wilson, Dalziel J and Zhang, Zheshen and others},
  journal={Communications Physics},
  volume={6},
  number={1},
  pages={237},
  year={2023},
  url={https://www.nature.com/articles/s41566-023-01178-0.pdf},
  publisher={Nature Publishing Group UK London}
}

@article{afek2022coherent,
  title={Coherent scattering of low mass dark matter from optically trapped sensors},
  author={Afek, Gadi and Carney, Daniel and Moore, David C},
  journal={Physical Review Letters},
  volume={128},
  number={10},
  pages={101301},
  year={2022},
  url={https://journals.aps.org/prl/abstract/10.1103/PhysRevLett.128.101301},
  publisher={APS}
}

@article{Sun2025,
  title = {Inner non-{H}ermitian skin effect on the {B}ethe lattice},
  author = {Sun, Junsong and Li, Chang-An and Li, Peilin and Feng, Shiping and Guo, Huaiming},
  journal = {Phys. Rev. B},
  volume = {111},
  issue = {7},
  pages = {075120},
  numpages = {11},
  year = {2025},
  month = {Feb},
  publisher = {American Physical Society},
  doi = {10.1103/PhysRevB.111.075120},
  url = {https://link.aps.org/doi/10.1103/PhysRevB.111.075120}
}

@article{wang2021generating,
  title={Generating arbitrary topological windings of a non-{H}ermitian band},
  author={Wang, Kai and Dutt, Avik and Yang, Ki Youl and Wojcik, Casey C and Vu{\v{c}}kovi{\'c}, Jelena and Fan, Shanhui},
  journal={Science},
  volume={371},
  number={6535},
  pages={1240--1245},
  year={2021},
  url={https://www.science.org/doi/full/10.1126/science.abf6568},
  publisher={American Association for the Advancement of Science}
}

@article{yuan2020creating,
  title={Creating locally interacting Hamiltonians in the synthetic frequency dimension for photons},
  author={Yuan, Luqi and Dutt, Avik and Qin, Mingpu and Fan, Shanhui and Chen, Xianfeng},
  journal={Photonics Research},
  volume={8},
  number={9},
  pages={B8--B14},
  year={2020},
  url={https://opg.optica.org/prj/fulltext.cfm?uri=prj-8-9-B8},
  publisher={Chinese Laser Press and Optical Society of America}
}

@article{koh2025interacting,
  title={Interacting non-Hermitian edge and cluster bursts on a digital quantum processor},
  author={Koh, Jin Ming and Xue, Wen-Tan and Tai, Tommy and Koh, Dax Enshan and Lee, Ching Hua},
  journal={arXiv preprint arXiv:2503.14595},
  url={https://arxiv.org/abs/2503.14595},
  year={2025}
}

@article{zee1998a,
title = {A non-{H}ermitian particle in a disordered world},
journal = {Physica A: Statistical Mechanics and its Applications},
volume = {254},
number = {1},
pages = {300-316},
year = {1998},
issn = {0378-4371},
doi = {https://doi.org/10.1016/S0378-4371(98)00030-2},
url = {https://www.sciencedirect.com/science/article/pii/S0378437198000302},
author = {A. Zee}
}

@article{rangi2025interplay,
  title={Interplay of non-{H}ermitian skin effect and electronic correlations in the non-{H}ermitian {H}ubbard model via Real-space dynamical mean field theory},
  author={Rangi, Chakradhar and Moreno, Juana and Tam, Ka-Ming},
  journal={arXiv preprint arXiv:2507.19471},
url={https://arxiv.org/pdf/2507.19471},
  year={2025}
}

@article{Hayata2021,
  title = {Non-{H}ermitian {H}ubbard model without the sign problem},
  author = {Hayata, Tomoya and Yamamoto, Arata},
  journal = {Phys. Rev. B},
  volume = {104},
  issue = {12},
  pages = {125102},
  numpages = {5},
  year = {2021},
  month = {Sep},
  publisher = {American Physical Society},
  doi = {10.1103/PhysRevB.104.125102},
  url = {https://link.aps.org/doi/10.1103/PhysRevB.104.125102}
}

@article{Longhi2023,
author = {Longhi, Stefano},
title = {Spectral Structure and Doublon Dissociation in the Two-Particle Non-{H}ermitian {H}ubbard Model},
journal = {Annalen der Physik},
volume = {535},
number = {11},
pages = {2300291},
doi = {https://doi.org/10.1002/andp.202300291},
url = {https://onlinelibrary.wiley.com/doi/abs/10.1002/andp.202300291},
year = {2023}
}

@article{Suthar2022,
  title = {Non-{H}ermitian many-body localization with open boundaries},
  author = {Suthar, Kuldeep and Wang, Yi-Cheng and Huang, Yi-Ping and Jen, H. H. and You, Jhih-Shih},
  journal = {Phys. Rev. B},
  volume = {106},
  issue = {6},
  pages = {064208},
  numpages = {11},
  year = {2022},
  month = {Aug},
  publisher = {American Physical Society},
  doi = {10.1103/PhysRevB.106.064208},
  url = {https://link.aps.org/doi/10.1103/PhysRevB.106.064208}
}

@article{hatano1998non,
  title={Non-{H}ermitian delocalization and eigenfunctions},
  author={Hatano, Naomichi and Nelson, David R},
  journal={Physical Review B},
  volume={58},
  number={13},
  pages={8384},
  year={1998},
url={https://journals.aps.org/prb/pdf/10.1103/PhysRevB.58.8384},
  publisher={APS}
}

@unpublished{ledoussalunpub,
  author = {P. Le Doussal},
  title  = {unpublished}
}

@article{hatano1996localization,
  title={Localization transitions in non-{H}ermitian quantum mechanics},
  author={Hatano, Naomichi and Nelson, David R},
  journal={Physical Review Letters},
  volume={77},
  number={3},
  pages={570},
  year={1996},
url={https://journals.aps.org/prl/pdf/10.1103/PhysRevLett.77.570},
  publisher={APS}
}

@article{bender2007making,
  title={Making sense of non-{H}ermitian {H}amiltonians},
  author={Bender, Carl M},
  journal={Reports on Progress in Physics},
  volume={70},
  number={6},
  pages={947},
  year={2007},
url={https://iopscience.iop.org/article/10.1088/0034-4885/70/6/R03/pdf},
  publisher={IOP Publishing}
}

@article{bergholtz2021exceptional,
  title={Exceptional topology of non-{H}ermitian systems},
  author={Bergholtz, Emil J and Budich, Jan Carl and Kunst, Flore K},
  journal={Reviews of Modern Physics},
  volume={93},
  number={1},
  pages={015005},
  year={2021},
url={https://journals.aps.org/rmp/abstract/10.1103/RevModPhys.93.015005},
  publisher={APS}
}

@article{rotter2009non,
  title={A non-{H}ermitian {H}amilton operator and the physics of open quantum systems},
  author={Rotter, Ingrid},
  journal={Journal of Physics A: Mathematical and Theoretical},
  volume={42},
  number={15},
  pages={153001},
  year={2009},
url={https://iopscience.iop.org/article/10.1088/1751-8113/42/15/153001/pdf},
  publisher={IOP Publishing}
}

@article{ding2022non,
  title={Non-{H}ermitian topology and exceptional-point geometries},
  author={Ding, Kun and Fang, Chen and Ma, Guancong},
  journal={Nature Reviews Physics},
  volume={4},
  number={12},
  pages={745--760},
  year={2022},
url={https://www.nature.com/articles/s42254-022-00516-5.pdf},
  publisher={Nature Publishing Group UK London}
}

@article{ghatak2019new,
  title={New topological invariants in non-{H}ermitian systems},
  author={Ghatak, Ananya and Das, Tanmoy},
  journal={Journal of Physics: Condensed Matter},
  volume={31},
  number={26},
  pages={263001},
  year={2019},
url={https://iopscience.iop.org/article/10.1088/1361-648x/ab11b3/pdf},
  publisher={IOP Publishing}
}

@article{zhang2022review,
  title={A review on non-{H}ermitian skin effect},
  author={Zhang, Xiujuan and Zhang, Tian and Lu, Ming-Hui and Chen, Yan-Feng},
  journal={Advances in Physics: X},
  volume={7},
  number={1},
  pages={2109431},
  year={2022},
url={https://www.tandfonline.com/doi/pdf/10.1080/23746149.2022.2109431},
  publisher={Taylor \& Francis}
}

@article{el2018non,
  title={Non-{H}ermitian physics and {PT} symmetry},
  author={El-Ganainy, Ramy and Makris, Konstantinos G and Khajavikhan, Mercedeh and Musslimani, Ziad H and Rotter, Stefan and Christodoulides, Demetrios N},
  journal={Nature Physics},
  volume={14},
  number={1},
  pages={11--19},
  year={2018},
url={https://www.nature.com/articles/nphys4323.pdf},
  publisher={Nature Publishing Group UK London}
}

@article{Gong2018,
  title = {Topological Phases of Non-{H}ermitian Systems},
  author = {Gong, Zongping and Ashida, Yuto and Kawabata, Kohei and Takasan, Kazuaki and Higashikawa, Sho and Ueda, Masahito},
  journal = {Phys. Rev. X},
  volume = {8},
  issue = {3},
  pages = {031079},
  numpages = {33},
  year = {2018},
  month = {Sep},
  publisher = {American Physical Society},
  doi = {10.1103/PhysRevX.8.031079},
  url = {https://link.aps.org/doi/10.1103/PhysRevX.8.031079}
}

@article{Borgnia2020,
  title = {Non-Hermitian Boundary Modes and Topology},
  author = {Borgnia, Dan S. and Kruchkov, Alex Jura and Slager, Robert-Jan},
  journal = {Phys. Rev. Lett.},
  volume = {124},
  issue = {5},
  pages = {056802},
  numpages = {6},
  year = {2020},
  month = {Feb},
  publisher = {American Physical Society},
  doi = {10.1103/PhysRevLett.124.056802},
  url = {https://link.aps.org/doi/10.1103/PhysRevLett.124.056802}
}

@article{Zhang2020,
  title = {Correspondence between Winding Numbers and Skin Modes in Non-{H}ermitian Systems},
  author = {Zhang, Kai and Yang, Zhesen and Fang, Chen},
  journal = {Phys. Rev. Lett.},
  volume = {125},
  issue = {12},
  pages = {126402},
  numpages = {6},
  year = {2020},
  month = {Sep},
  publisher = {American Physical Society},
  doi = {10.1103/PhysRevLett.125.126402},
  url = {https://link.aps.org/doi/10.1103/PhysRevLett.125.126402}
}

@article{Okuma2020,
  title = {Topological Origin of Non-{H}ermitian Skin Effects},
  author = {Okuma, Nobuyuki and Kawabata, Kohei and Shiozaki, Ken and Sato, Masatoshi},
  journal = {Phys. Rev. Lett.},
  volume = {124},
  issue = {8},
  pages = {086801},
  numpages = {7},
  year = {2020},
  month = {Feb},
  publisher = {American Physical Society},
  doi = {10.1103/PhysRevLett.124.086801},
  url = {https://link.aps.org/doi/10.1103/PhysRevLett.124.086801}
}

@article{Kawabata2022,
  title = {Many-body topology of non-{H}ermitian systems},
  author = {Kawabata, Kohei and Shiozaki, Ken and Ryu, Shinsei},
  journal = {Phys. Rev. B},
  volume = {105},
  issue = {16},
  pages = {165137},
  numpages = {11},
  year = {2022},
  month = {Apr},
  publisher = {American Physical Society},
  doi = {10.1103/PhysRevB.105.165137},
  url = {https://link.aps.org/doi/10.1103/PhysRevB.105.165137}
}

@article{ashida2020non,
  title={Non-{H}ermitian physics},
  author={Ashida, Yuto and Gong, Zongping and Ueda, Masahito},
  journal={Advances in Physics},
  volume={69},
  number={3},
  pages={249--435},
  year={2020},
url={https://www.tandfonline.com/doi/pdf/10.1080/00018732.2021.1876991},
  publisher={Taylor \& Francis}
}

@article{hatano1997vortex,
  title={Vortex pinning and non-{H}ermitian quantum mechanics},
  author={Hatano, Naomichi and Nelson, David R},
  journal={Physical Review B},
  volume={56},
  number={14},
  pages={8651},
  year={1997},
url={https://journals.aps.org/prb/pdf/10.1103/PhysRevB.56.8651},
  publisher={APS}
}

@article{feinberg1999spectral,
  title={Spectral curves of non-{H}ermitian {H}amiltonians},
  author={Feinberg, Joshua and Zee, A},
  journal={Nuclear Physics B},
  volume={552},
  number={3},
  pages={599--623},
  year={1999},
url={https://www.sciencedirect.com/science/article/pii/S0550321399002461},
  publisher={Elsevier}
}

@article{feinberg1999non,
  title={Non-{H}ermitian localization and delocalization},
  author={Feinberg, Joshua and Zee, A},
  journal={Physical Review E},
  volume={59},
  number={6},
  pages={6433},
  year={1999},
url={https://journals.aps.org/pre/pdf/10.1103/PhysRevE.59.6433},
  publisher={APS}
}

@article{song2019non,
  title={Non-{H}ermitian skin effect and chiral damping in open quantum systems},
  author={Song, Fei and Yao, Shunyu and Wang, Zhong},
  journal={Physical Review Letters},
  volume={123},
  number={17},
  pages={170401},
  year={2019},
url={https://journals.aps.org/prl/pdf/10.1103/PhysRevLett.123.170401},
  publisher={APS}
}

@article{shen2018topological,
  title={Topological band theory for non-{H}ermitian {H}amiltonians},
  author={Shen, Huitao and Zhen, Bo and Fu, Liang},
  journal={Physical Review Letters},
  volume={120},
  number={14},
  pages={146402},
  year={2018},
url={https://journals.aps.org/prl/pdf/10.1103/PhysRevLett.120.146402},
  publisher={APS}
}

@article{Kawabata2019,
  title={Symmetry and topology in non-{H}ermitian physics},
  author={Kawabata, Kohei and Shiozaki, Ken and Ueda, Masahito and Sato, Masatoshi},
  journal={Physical Review X},
  volume={9},
  number={4},
  pages={041015},
  year={2019},
url={https://journals.aps.org/prx/pdf/10.1103/PhysRevX.9.041015},
  publisher={APS}
}

@article{feng2017non,
  title={Non-{H}ermitian photonics based on parity--time symmetry},
  author={Feng, Liang and El-Ganainy, Ramy and Ge, Li},
  journal={Nature Photonics},
  volume={11},
  number={12},
  pages={752--762},
  year={2017},
url={https://www.nature.com/articles/s41566-017-0031-1},
  publisher={Nature Publishing Group}
}

@article{Lee2019,
  title = {Hybrid Higher-Order Skin-Topological Modes in Nonreciprocal Systems},
  author = {Lee, Ching Hua and Li, Linhu and Gong, Jiangbin},
  journal = {Phys. Rev. Lett.},
  volume = {123},
  issue = {1},
  pages = {016805},
  numpages = {6},
  year = {2019},
  month = {Jul},
  publisher = {American Physical Society},
  doi = {10.1103/PhysRevLett.123.016805},
  url = {https://link.aps.org/doi/10.1103/PhysRevLett.123.016805}
}

@article{li2020critical,
  title={Critical non-{H}ermitian skin effect},
  author={Li, Linhu and Lee, Ching Hua and Mu, Sen and Gong, Jiangbin},
  journal={Nature communications},
  volume={11},
  number={1},
  pages={5491},
  year={2020},
  url={https://www.nature.com/articles/s41467-020-18917-4.pdf},
  publisher={Nature Publishing Group UK London}
}

@ARTICLE{Peierls,
       author = {{Peierls}, R.},
        title = "{Zur Theorie des Diamagnetismus von Leitungselektronen}",
      journal = {Zeitschrift fur Physik},
         year = 1933,
        month = nov,
       volume = {80},
       number = {11-12},
        pages = {763-791},
          doi = {10.1007/BF01342591},
       adsurl = {https://ui.adsabs.harvard.edu/abs/1933ZPhy...80..763P}
}

@article{Luttinger,
  title = {The Effect of a Magnetic Field on Electrons in a Periodic Potential},
  author = {Luttinger, J. M.},
  journal = {Phys. Rev.},
  volume = {84},
  issue = {4},
  pages = {814--817},
  numpages = {0},
  year = {1951},
  month = {Nov},
  publisher = {American Physical Society},
  doi = {10.1103/PhysRev.84.814},
  url = {https://link.aps.org/doi/10.1103/PhysRev.84.814}
}

@article{Lee2016Ano,
  title = {Anomalous Edge State in a Non-{H}ermitian Lattice},
  author = {Lee, Tony E.},
  journal = {Phys. Rev. Lett.},
  volume = {116},
  issue = {13},
  pages = {133903},
  numpages = {5},
  year = {2016},
  month = {Apr},
  publisher = {American Physical Society},
  doi = {10.1103/PhysRevLett.116.133903},
  url = {https://link.aps.org/doi/10.1103/PhysRevLett.116.133903}
}

@article{Xiong_2018,
doi = {10.1088/2399-6528/aab64a},
url = {https://doi.org/10.1088/2399-6528/aab64a},
year = {2018},
month = {mar},
publisher = {IOP Publishing},
volume = {2},
number = {3},
pages = {035043},
author = {Xiong, Ye},
title = {Why does bulk boundary correspondence fail in some non-{H}ermitian topological models},
journal = {Journal of Physics Communications},
}

@article{Daley2014,
author = {Andrew J. Daley},
title = {Quantum trajectories and open many-body quantum systems},
journal = {Advances in Physics},
volume = {63},
number = {2},
pages = {77--149},
year = {2014},
publisher = {Taylor \& Francis},
doi = {10.1080/00018732.2014.933502}}

@article{dalibard1992wave,
  title = {Wave-function approach to dissipative processes in quantum optics},
  author = {Dalibard, Jean and Castin, Yvan and M\o{}lmer, Klaus},
  journal = {Phys. Rev. Lett.},
  volume = {68},
  issue = {5},
  pages = {580--583},
  numpages = {0},
  year = {1992},
  month = {Feb},
  publisher = {American Physical Society},
  doi = {10.1103/PhysRevLett.68.580},
  url = {https://link.aps.org/doi/10.1103/PhysRevLett.68.580}
}

@article{molmer1993monte,
author = {Klaus M{\o}lmer and Yvan Castin and Jean Dalibard},
journal = {J. Opt. Soc. Am. B},
number = {3},
pages = {524--538},
publisher = {Optica Publishing Group},
title = {Monte Carlo wave-function method in quantum optics},
volume = {10},
month = {Mar},
year = {1993},
url = {https://opg.optica.org/josab/abstract.cfm?URI=josab-10-3-524},
doi = {10.1364/JOSAB.10.000524},
}

@article{dum1992spe,
  title = {Monte Carlo simulation of the atomic master equation for spontaneous emission},
  author = {Dum, R. and Zoller, P. and Ritsch, H.},
  journal = {Phys. Rev. A},
  volume = {45},
  issue = {7},
  pages = {4879--4887},
  numpages = {0},
  year = {1992},
  month = {Apr},
  publisher = {American Physical Society},
  doi = {10.1103/PhysRevA.45.4879},
  url = {https://link.aps.org/doi/10.1103/PhysRevA.45.4879}
}

@article{dum1992vtsr,
  title = {Monte Carlo simulation of master equations in quantum optics for vacuum, thermal, and squeezed reservoirs},
  author = {Dum, R. and Parkins, A. S. and Zoller, P. and Gardiner, C. W.},
  journal = {Phys. Rev. A},
  volume = {46},
  issue = {7},
  pages = {4382--4396},
  numpages = {0},
  year = {1992},
  month = {Oct},
  publisher = {American Physical Society},
  doi = {10.1103/PhysRevA.46.4382},
  url = {https://link.aps.org/doi/10.1103/PhysRevA.46.4382}
}

@article{Yu2024,
  title = {Non-Hermitian Strongly Interacting {D}irac Fermions},
  author = {Yu, Xue-Jia and Pan, Zhiming and Xu, Limei and Li, Zi-Xiang},
  journal = {Phys. Rev. Lett.},
  volume = {132},
  issue = {11},
  pages = {116503},
  numpages = {8},
  year = {2024},
  month = {Mar},
  publisher = {American Physical Society},
  doi = {10.1103/PhysRevLett.132.116503},
  url = {https://link.aps.org/doi/10.1103/PhysRevLett.132.116503}
}

@article{Mu2020,
  title = {Emergent {F}ermi surface in a many-body non-{H}ermitian fermionic chain},
  author = {Mu, Sen and Lee, Ching Hua and Li, Linhu and Gong, Jiangbin},
  journal = {Phys. Rev. B},
  volume = {102},
  issue = {8},
  pages = {081115},
  numpages = {8},
  year = {2020},
  month = {Aug},
  publisher = {American Physical Society},
  doi = {10.1103/PhysRevB.102.081115},
  url = {https://link.aps.org/doi/10.1103/PhysRevB.102.081115}
}

@article{Lee2020,
  title = {Many-body approach to non-Hermitian physics in fermionic systems},
  author = {Lee, Eunwoo and Lee, Hyunjik and Yang, Bohm-Jung},
  journal = {Phys. Rev. B},
  volume = {101},
  issue = {12},
  pages = {121109},
  numpages = {6},
  year = {2020},
  month = {Mar},
  publisher = {American Physical Society},
  doi = {10.1103/PhysRevB.101.121109},
  url = {https://link.aps.org/doi/10.1103/PhysRevB.101.121109}
}

@article{Wang2023,
  title = {Non-{H}ermitian {H}aldane-{H}ubbard model: {E}ffective description of one- and two-body dissipation},
  author = {Wang, Can and Yi, Tian-Cheng and Li, Jian and Mondaini, Rubem},
  journal = {Phys. Rev. B},
  volume = {108},
  issue = {8},
  pages = {085134},
  numpages = {9},
  year = {2023},
  month = {Aug},
  publisher = {American Physical Society},
  doi = {10.1103/PhysRevB.108.085134},
  url = {https://link.aps.org/doi/10.1103/PhysRevB.108.085134}
}

@article{Yi2025,
  title = {Topological and {M}ott phases driven by gain and loss in a correlated {C}hern insulator},
  author = {Yi, Tian-Cheng and Mondaini, Rubem},
  journal = {Phys. Rev. B},
  volume = {112},
  issue = {4},
  pages = {045120},
  numpages = {10},
  year = {2025},
  month = {Jul},
  publisher = {American Physical Society},
  doi = {10.1103/1xnb-s8w6},
  url = {https://link.aps.org/doi/10.1103/1xnb-s8w6}
}

@article{Orito2023,
  title = {Entanglement dynamics in the many-body {Hatano-Nelson} model},
  author = {Orito, Takahiro and Imura, Ken-Ichiro},
  journal = {Phys. Rev. B},
  volume = {108},
  issue = {21},
  pages = {214308},
  numpages = {18},
  year = {2023},
  month = {Dec},
  publisher = {American Physical Society},
  doi = {10.1103/PhysRevB.108.214308},
  url = {https://link.aps.org/doi/10.1103/PhysRevB.108.214308}
}

@article{Zhang2022,
  title = {Symmetry breaking and spectral structure of the interacting {H}atano-{N}elson model},
  author = {Zhang, Song-Bo and Denner, M. Michael and Bzdu\ifmmode \check{s}\else \v{s}\fi{}ek, Tom\'a\ifmmode \check{s}\else \v{s}\fi{} and Sentef, Michael A. and Neupert, Titus},
  journal = {Phys. Rev. B},
  volume = {106},
  issue = {12},
  pages = {L121102},
  numpages = {8},
  year = {2022},
  month = {Sep},
  publisher = {American Physical Society},
  doi = {10.1103/PhysRevB.106.L121102},
  url = {https://link.aps.org/doi/10.1103/PhysRevB.106.L121102}
}

@article{Faugno2022,
  title = {Interaction-Induced Non-{H}ermitian Topological Phases from a Dynamical Gauge Field},
  author = {Faugno, W. N. and Ozawa, Tomoki},
  journal = {Phys. Rev. Lett.},
  volume = {129},
  issue = {18},
  pages = {180401},
  numpages = {6},
  year = {2022},
  month = {Oct},
  publisher = {American Physical Society},
  doi = {10.1103/PhysRevLett.129.180401},
  url = {https://link.aps.org/doi/10.1103/PhysRevLett.129.180401}
}

@article{Dora2022,
  title = {Full counting statistics in the many-body {Hatano-Nelson} model},
  author = {D\'ora, Bal\'azs and Moca, C\ifmmode \u{a}\else \u{a}\fi{}t\ifmmode \u{a}\else \u{a}\fi{}lin Pa\ifmmode \mbox{\c{s}}\else \c{s}\fi{}cu},
  journal = {Phys. Rev. B},
  volume = {106},
  issue = {23},
  pages = {235125},
  numpages = {6},
  year = {2022},
  month = {Dec},
  publisher = {American Physical Society},
  doi = {10.1103/PhysRevB.106.235125},
  url = {https://link.aps.org/doi/10.1103/PhysRevB.106.235125}
}

@article{Dupays2025,
  title = {Slow approach to adiabaticity in many-body non-{H}ermitian systems: The {Hatano-Nelson} model},
  author = {Dupays, L\'eonce and del Campo, Adolfo and D\'ora, Bal\'azs},
  journal = {Phys. Rev. B},
  volume = {111},
  issue = {4},
  pages = {045130},
  numpages = {7},
  year = {2025},
  month = {Jan},
  publisher = {American Physical Society},
  doi = {10.1103/PhysRevB.111.045130},
  url = {https://link.aps.org/doi/10.1103/PhysRevB.111.045130}
}

@article{Alsallom2022,
  title = {Fate of the non-{H}ermitian skin effect in many-body fermionic systems},
  author = {Alsallom, Faisal and Herviou, Lo\"{\i}c and Yazyev, Oleg V. and Brzezi\ifmmode \acute{n}\else \'{n}\fi{}ska, Marta},
  journal = {Phys. Rev. Res.},
  volume = {4},
  issue = {3},
  pages = {033122},
  numpages = {11},
  year = {2022},
  month = {Aug},
  publisher = {American Physical Society},
  doi = {10.1103/PhysRevResearch.4.033122},
  url = {https://link.aps.org/doi/10.1103/PhysRevResearch.4.033122}
}

@article{Pu2026,
  title = {Acoustic Bound Pairs under Nonreciprocal Two-Body Interactions},
  author = {Pu, Zhenhang and Xi, Yuxiang and Tang, Yugan and Lu, Jiuyang and Deng, Weiyin and Cheng, Hua and Ke, Manzhu and Chen, Shuqi and Liu, Zhengyou},
  journal = {Phys. Rev. Lett.},
  volume = {136},
  issue = {14},
  pages = {146603},
  numpages = {7},
  year = {2026},
  month = {Apr},
  publisher = {American Physical Society},
  doi = {10.1103/xr3h-j9pv},
  url = {https://link.aps.org/doi/10.1103/xr3h-j9pv}
}

@misc{zenodo,
  title        = {{Data and figures for ``The Two Orbital, Interacting Hatano-Nelson Model''}},
  year         = {2026},
  publisher    = {Zenodo},
  doi          = {10.5281/zenodo.19599761},
  url          = {https://doi.org/10.5281/zenodo.19599761},
  note         = {{Z}enodo repository containing all data and figures}
}

\end{document}